\documentclass[preprints,article,accept,moreauthors,pdftex]{Definitions/mdpi} 
\usepackage{amsmath,amssymb,amsfonts}
\usepackage{algorithmic}
\usepackage{graphicx}
\usepackage{subcaption}
\usepackage{textcomp}
\usepackage{xcolor}
\usepackage{tikz}
\usepackage{boondox-cal}
\DeclareMathAlphabet\mathbfcal{OMS}{cmsy}{b}{n}
\def\BibTeX{{\rm B\kern-.05em{\sc i\kern-.025em b}\kern-.08em
		T\kern-.1667em\lower.7ex\hbox{E}\kern-.125emX}}
\firstpage{1} 
\pubvolume{xx}
\issuenum{1}
\articlenumber{5}
\pubyear{2020}
\copyrightyear{2020}
\history{Received: date; Accepted: date; Published: date}

\Title{Estimating Conditional Transfer Entropy in Time Series using Mutual Information and Non-linear Prediction}

\Author{Payam Shahsavari Baboukani $^{1,*}$, Carina Graversen $^{2}$, Emina Alickovic $^{2,3}$ and Jan Østergaard $^{1}$}

\AuthorNames{Payam Shahsavari Baboukani, Carina Graversen, Emina Alickovic and Jan Østergaard}

\address{%
$^{1}$ \quad Department of Electronic Systems, Aalborg University, Denmark; Email: jo@es.aau.dk\\
$^{2}$ \quad Eriksholm Research Centre, Oticon A/S, Denmark; E-Mails: cagr@eriksholm.com and eali@eriksholm.com\\
$^{3}$ \quad Department of Electrical Engineering, Linköping University, Linköping, Sweden}
\corres{Correspondence: pasba@es.aau.dk}

\conference{\cite{My}} 

\abstract{We propose a new estimator to measure directed dependencies in time series.
The dimensionality of data is first reduced using a new non-uniform embedding technique, where the variables are ranked according to a weighted sum of the amount of new information and improvement of the prediction accuracy provided by the variables. 
Then, using a greedy approach, the most informative subsets are selected in an iterative way. The algorithm terminates, when the highest ranked variable is not able to significantly improve the accuracy of the prediction as compared to that obtained using the existing selected subsets. 
In a simulation study, we compare our estimator to existing state-of-the-art methods at different data lengths and directed dependencies strengths.
It is demonstrated that the proposed estimator has a significantly higher accuracy than that of existing methods, especially for the difficult case, where the data is highly correlated and coupled. Moreover, we show its false detection of directed dependencies due to instantaneous couplings effect is lower than that of existing measures. We also show applicability of the proposed estimator on real intracranial electroencephalography data.}


\begin{document}


\section{Introduction}
Real-world interconnected technological systems such as car traffic and distributed power grids as well as biological systems such as the human brain can be represented in terms of complex dynamical systems that contain subsystems. Characterizing the subsystems and their interdependencies can help understanding the overall system behavior on a local and global scale. For example, different regions of the brain such as the cortices can be considered as subsystems. An assessment of the interaction between the cortices may provide insights into how the brain functions \cite{omidvarnia2013measuring}. In order to identify the interactions, several time series analyses methods ranging from information theoretical to signal processing  approaches have been proposed in the literature \cite{cover2012elements,baboukani2019novel,schreiber2000measuring}. In particular, the directional methods have gained increasing attention because, unlike symmetric measures such as mutual information \cite{cover2012elements} and phase synchronization \cite{baboukani2019novel,baboukani2017classifying}, directional measures are generally able to assess the direction in addition to the strength of the interactions between subsystems \cite{schreiber2000measuring,Gen2018,faes2017multiscale,derpich2013fundamental,massey1990causality}.

A popular approach used in the literature to assess directed dependencies uses Wiener’s definition, which is based on the concept of prediction \cite{wiener1956theory}. According to the Wiener’s definition, if the prediction of the future value of a time series $X_t$ from its own past values can be improved by incorporating past values of another time series $Y_t$, then there are causal dependencies from $Y_t$ to $X_t$ \cite{wiener1956theory}. Although the term “causal” was used in Wiener’s definition, it has been shown that measures quantifying the Wiener’s definition over- or under-estimate the causal effect in certain cases \cite{james2016information,lizier2010differentiating}. In this paper, we use the term “directed dependencies” to refer to the property of time series or processes satisfying Wiener’s definition.

Schreiber \cite{schreiber2000measuring} formalized directed dependencies by using the concept of conditional mutual information (CMI) and proposed a new measure called transfer entropy (TE). TE does not depend on any model in its formulation, which makes this method able to assess both linear and non-linear interactions \cite{montalto2014mute}. Additionally, estimating TE by using the combination of data-efficient and model-free estimators like Kraskov-St\"{o}gbauer-Grassberger (KSG) \cite{kraskov2004estimating}, and uniform embedding state space reconstruction schemes \cite{lindner2011trentool,wibral2013measuring} has increased the popularity of TE. TE has been used for quantifying directed dependencies between joint processes in neuro-physiological \cite{lindner2011trentool,wibral2013measuring} and economical \cite{bossomaier2016introduction} applications.

As an example, assume that we are interested in measuring TE between processes which for example, represent sensor measurement data from different regions of the brain, e.g., multi-channel electroencephalography (EEG) data. The recorded EEG data is spatially auto-correlated due to the phenomenon known as the volume conduction effect in neuro-physiological time series \cite{ruiz2019computational}. The spatial auto-correlation in such data can lead to overestimate in the estimated TE and eventually lead to false positives detection of TE. A possible approach to reduce such effect is to use conditional version of TE \cite{faes2016information,mehta2017directional}, which referred to as conditional transfer entropy (CTE). 

It is preferred to condition out all other variables in the network to ensure that the obtained CTE values reflect the true directed dependencies from an individual source to the target. On the other hand, the more variables we include in the conditioning, the higher the dimension of the problem becomes and the less accurate CTE estimators are, since we only have access to limited number of realizations. Considering the fact that we are interested in estimating directed dependencies and we need to condition out past variables related to the remaining variables, the dimension of the conditioning process increases even more and reliable estimation of CTE in multi-channel data (such as EEG data) by using the classical uniform embedding technique is limited by the so-called “curse of dimensionality” problem \cite{montalto2014mute,zhang2018low,xiong2017entropy,jia2019detecting}.

Non-uniform embedding (NUE) approaches reconstruct the past of the system with respect to a target variable by selecting the most relevant past and thereby decreases the dimensionality  \cite{montalto2014mute,faes2016information,xiong2017entropy,kugiumtzis2013direct,olejarczyk2017comparison,novelli2019large}. The information theoretical-based NUE algorithm proposed in \cite{montalto2014mute} is a greedy strategy, which uses CMI for selecting the most informative candidates. The authors in \cite{montalto2014mute} showed a significant improvement of NUE over uniform embedding. The author in  \cite{zhang2018low} stated that as the iteration of the NUE algorithm increases and more variables are selected, estimation of the higher dimensional CMI may become less accurate. The author in \cite{zhang2018low}, then suggested to use a low-dimensional approximation (LA) of the CMI, and proposed a new NUE algorithm. 

Adding more variables in the conditioning process decreases accuracy of the CTE estimator. The key problem is therefore how to decide whether we should include more variables, or terminate the algorithm. The existing NUE algorithms terminate if they fulfill a termination criterion defined by a bootstrap statistical-based test \cite{montalto2014mute,jia2019detecting,zhang2018low,novelli2019large}. The bootstrap test is used to approximate a confidence bound (or a critical value) by which the NUE algorithm is terminated. A higher bootstrap size, up to a threshold, generally leads to better approximation of the confidence bound \cite{may2008non}, which can further influence the accuracy of the NUE algorithms. A bootstrap size of at most 100 is generally used in the literature \cite{montalto2014mute,faes2016information,xiong2017entropy,zhang2018low} due to computational complexity reasons. It has been shown that using an alternative to the bootstrap-based termination criterion can improve the accuracy and computational efficiency of the greedy algorithms \cite{may2008non,li2015improved}. For example, the Akaike information criterion (AIC) and kernel density estimation (KDE)-based regression was proposed in \cite{may2008non} as an alternative to bootstrap methods for input variable selection techniques

In the present study, inspired by \cite{may2008non}, we propose an alternative approach to the bootstrap-based termination criterion used in the existing NUE algorithms. Specifically, to aid in making the decision of whether to include a variable or terminate the algorithm, we propose to measure the relevance of the new candidate variable by assessing the effect of it on the accuracy of the non-linear prediction of the target variable. The non-linear prediction is based on nearest neighbor (NN)-based regression \cite{altman1992introduction}. We show that it is also advantageous to use the non-linear prediction strategy for selecting the pool of candidates in the first place. We then introduce a new NUE algorithm which uses a weighted combination of CMI and the accuracy of the non-linear prediction for selection of candidates and present the new termination criterion for stopping the algorithm. Finally, we demonstrate that our proposed NUE procedure is more accurate, than the existing NUE algorithms on both synthetic and real-world data. 

The effect of instantaneous coupling (IC) on the NUE algorithms will also be investigated. IC can occur due to simultaneous (zero lag) information sharing like source mixing as a result of volume conduction in EEG signals \cite{faes2016information,faes2013compensated} and may lead to spurious detection of TE or CTE. 

The remainder of this paper is structured as follows. In section 2, the necessary background on CTE and the existing NUE algorithms will be briefly reviewed. Then, the proposed termination criterion and NUE procedure will be introduced in Section 3 and 4, respectively. This is followed by the description of our simulation study in Section 5, which is based on Henon maps and non-linear autoregressive (AR) models. The results of applying the proposed NUE algorithm on real EEG data will be reported in Section 6. Section 7 will discuss the results. The same section will also conclude the paper.
\section{Background}
\subsection{conditional Transfer entropy}
Let us consider a complex system which consists of $L$ interacting subsystems. We assume that we are interested in assessing the directed dependencies between subsystems $\mathcal{X}$ and $\mathcal{Y}$. Let stationary stochastic processes $X=(X_1, X_2, \dots, X_N)$ and $Y=(Y_1, Y_2, \dots, Y_N)$ describe the state visited by the subsystem $\mathcal{X}$ and $\mathcal{Y}$ over time, respectively. We denote $X_n\in\mathbb{R}$ and $Y_n\in\mathbb{R}$ as stochastic variables obtained by sampling the processes $X$ and $Y$ at the present time $n$, respectively. Furthermore, we denote the past of $X$ up till $X_{n-1}$ by a random vector $X_n^-=[X_{n-1},X_{n-2},...]$. TE from $X$ to $Y$ is then defined as \cite{schreiber2000measuring}
\begin{figure}[b]
	\vspace{0.2cm}
	\includegraphics[width=7cm,trim={0 0.5cm 0 0.5cm},clip]{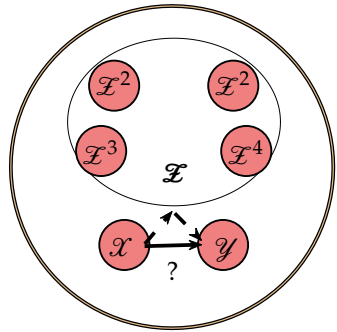}
	\put(-100,65){$\boldsymbol{\mathcal{Z}}$}
	\put(-130,72){$\mathcal{Z}^3$}
	\put(-75,72){$\mathcal{Z}^4$}
	\put(-80,98){$\mathcal{Z}^2$}
	\put(-125,98){$\mathcal{Z}^2$}
	\put(-80,37){$\mathcal{Y}$}
	\put(-120,37){$\mathcal{X}$}
	\put(-98,28){?}
	\label{Net-Indi}
	\centering
	\label{Net_Com}
	\caption{An example of $L=6$ nodes network where indirect paths through the remaining channels $\boldsymbol{\mathcal{Z}}$ may cause a falsely (dashed line) detected directed dependency (solid line) from $X$ to $Y$.}
	\label{Net}
\end{figure}

\begin{equation}
\operatorname{TE}(X\rightarrow Y)\triangleq I(Y_n;X_n^-\mid Y_n^-),
\label{TE}
\end{equation}
where $I(\,.\,;\,.\mid.)$ is CMI. However, in a complex network, it is not guaranteed that \eqref{TE} only describes the directed dependencies from $X$ to $Y$. For example, there could be a third process, say $Z$, through  which shared information is mediated to $X$ and $Y$. In this case, the shared information will lead to an increase in TE. To reduce the effect of common information being shared through other process, it has been suggested to use CTE \cite{montalto2014mute,faes2016information}. Let us consider the $L=6$ nodes network in Figure \ref{Net}, where we are interested in assessing the directed dependencies from node $\mathcal{X}$ to $\mathcal{Y}$ and which is not due to indirect paths through the remaining nodes $\boldsymbol{\mathcal{Z}}=\{\mathcal{Z}^1,\mathcal{Z}^2,\mathcal{Z}^3,\mathcal{Z}^4\}$. We denote $Z^i=(Z^i_1,Z^i_2,\dots,Z^i_N)$ as a stochastic process describes the state visited by $\mathcal{Z}^i$ and   $\boldsymbol{Z}=[Z^1,Z^2,...Z^{4}]$ as a $4$-variate stochastic process which describes state visited by $\boldsymbol{\mathcal{Z}}$ over time. CTE from an individual source $X$ to the target $Y$ excluding information from $\boldsymbol{Z}$ is then defined as 
  
\begin{equation}
\operatorname{CTE}(X\rightarrow Y|\boldsymbol{Z})\triangleq I(Y_n;X_n^-\mid Y_n^-,\boldsymbol{Z}_n^-),
\label{PTE}
\end{equation}
where $\boldsymbol{Z}_n^-=[\boldsymbol{Z}_{n-1},\boldsymbol{Z}_{n-2},...]$ denotes the past of up $\boldsymbol{Z}$ till but not including $\boldsymbol{Z}_n$.

\subsection{Existing Non-uniform embedding algorithm}
Prior to estimating CTE in \eqref{PTE}, it is mandatory to approximate the possibly infinite-dimensional random vectors which represent the past of the processes. Let us denote the approximated past vector variable $X_n^-$ by $V_n^X$. The same notation applies to $V_n^Y$ and $V_n^{\boldsymbol{Z}}$. The basic idea behind reconstructing the past of the processes $X$, $Y$, and $\boldsymbol{Z}$ by assuming $Y$ as the target process is to form a low dimensional embedding vector $\mathcal{S}$ comprising the most informative past variables about the present state of the target $Y$. Traditionally, the past of the system is reconstructed by using the uniform embedding scheme in which each component of $\mathcal{S}$ is approximated separately. For example, $V_n^Y$ is approximated as, $V_n^Y=[Y_{n-m},Y_{n-2m},...,Y_{n-dm}]$ where $m$ and $d$ are the embedding delay and embedding dimension, respectively \cite{lindner2011trentool,montalto2014mute}. Then the $V_n^X$ and $V_n^{\boldsymbol{Z}}$ are estimated using the same approach and the final embedding vector $\mathcal{S}=[V_n^X,V_n^Y,V_n^{\textbf{Z}}]$ is formed and utilized to estimated CTE in \eqref{PTE}.

The uniform embedding scheme may lead to selection of redundant past variables and ignore relevant variables, as a result decrease the accuracy of the CTE estimation. This can limit applications in high dimensional data \cite{montalto2014mute,jia2019detecting,zhang2018low}. Alternatively, the NUE schemes try to select the most relevant and least redundant past variables and form a new embedding vector \cite{montalto2014mute,jia2019detecting,zhang2018low}.
\subsubsection{Bootstrap-based Non-uniform Embedding Algorithm}

 The NUE algorithm, as suggested in \cite{montalto2014mute}, can be described as follows:

\begin{enumerate}[leftmargin=*,labelsep=4.9mm]
	\item Choose embedding delay $d$ and embedding dimension $m$ and construct the candidate set $\mathcal{C}=[X_{n-m},...,X_{n-md},Y_{n-m},..Y_{n-md},\boldsymbol{Z}_{n-m},...,\textbf{Z}_{n-md}]$.
	\item	Initialize the algorithms by an empty set of the selected candidates $\mathcal{S}_n^0=\emptyset$.
	\item	Run a forward search to find the most informative candidate among the candidate set $\mathcal{C}$. This can be achieved by quantifying the amount of information that each candidate $W_n$ has about $Y_n$ which is not provided by the selected candidates from the last iteration  $\mathcal{S}_n^{k-1}$. To formalize this, at each iteration $k\geq1$, select the candidate $W_n^k$, such that CMI between $W_n$ and $Y_n$ conditioned on $\mathcal{S}_n^{k-1}$ is maximized
	\begin{equation}
	W_n^k=\underset{W_n\in \mathcal{C}\setminus \mathcal{S}_n^{k-1}}{\operatorname{argmax}}\operatorname{I}\left(Y_n;W_n\mid\mathcal{S}_n^{k-1}\right),
	\label{Select}
	\end{equation}	
	where $\mathcal{S}_n^{k-1}=\bigcup\limits_{i=0}^{k-1}W_n^i$ denotes the set of the selected candidates up till iteration $k-1$ and $\mathcal{C}\setminus \mathcal{S}_n^{k-1}$ denotes the remaining candidates in $\mathcal{C}$. We estimate the CMI given in \eqref{Select} by using the KSG approach \cite{kraskov2004estimating,montalto2014mute,faes2015estimating} in this study (cf. Appendix \ref{NN}).
	\item  Stop the iteration if the termination criterion is fulfilled and return $\mathcal{S}_n^{k-1}$ as the desired embedding vector.
\end{enumerate}

The flow chart of the NUE algorithm is shown in Figure \ref{NUEFlow}. After obtaining the embedding vector $\mathcal{S}_n^{k-1}$, CTE is estimated by using \eqref{PTE} in which case $[X_n^-,Y_n^-,\textbf{Z}_n^-]$ is replaced by $\mathcal{S}_n^{k-1}$ and $[Y_n^-,\textbf{Z}_n^-]$ is replaced by $\mathcal{S}_n^{k-1}$ excluding the past of $X_n$. CTE is written as the sum/difference of four differential entropies and is estimated by using KSG approach \footnote{In this paper, we use the KSG approach to estimate CTE and CMI. The KSG estimator is designed to estimates differential entropies. Therefore, we assumed that variables used in this paper are continuous.} \cite{kraskov2004estimating,montalto2014mute,faes2015estimating} (cf. Appendix \ref{NNPTE}). 

The existing NUE algorithm proposed in \cite{montalto2014mute} utilizes a bootstrap-based termination criterion. The goal of the bootstrap test in the NUE algorithm is to estimate an upper bound on the CMI between independently selected candidate $\widehat{W_n^k}$ and the target variable $\widehat{Y_n}$ given $\mathcal{S}_n^{k-1}$, $I(\widehat{W_n^k};\widehat{Y_n}|\mathcal{S}_n^{k-1})$. The estimation is accomplished by drawing 100 independent randomly shuffled realizations of $Y_n$ and $W_n^k$, estimating the CMI between the randomized $W_n^k$ and the randomized $Y_n$ given the original $\mathcal{S}_n^{k-1}$, and then finding the $95^{th}$ percentile $I^{95}$ of the generated distribution. The obtained value $I^{95}$ can be used as a critical value (at $5\%$ confidence level) of $I({W_n^k};{Y_n}|\mathcal{S}_n^{k-1})$ so that if $I({W_n^k};{Y_n}|\mathcal{S}_n^{k-1})>I^{95}$ then the candidate is included in the embedding vector and the algorithm continues to search for more candidates in iteration $k+1$. Otherwise, the termination criterion is fulfilled and the algorithm is ended and $\mathcal{S}_n^{k-1}$ is returned as the embedding vector.

\subsubsection{Low-dimensional Approximation-based Non-uniform Embedding Algorithm} 

\begin{figure}[t]
 	\centering
 	\includegraphics[width=10cm,trim={0 0.01cm 0 0.01cm},clip]{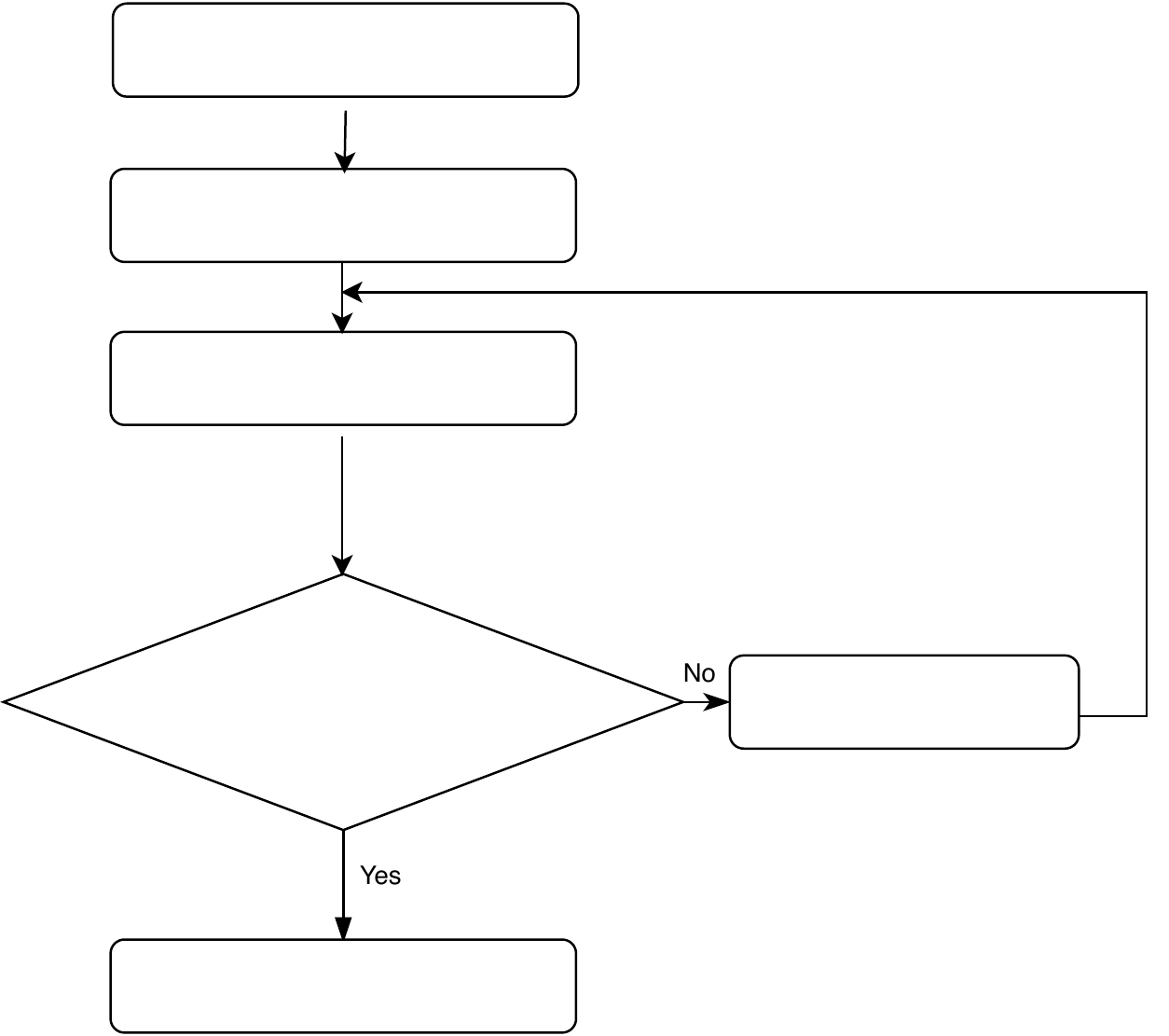}
 	\put(-240,238){Choose to $m$ and $d$}
 	\put(-240,198){Set $\mathcal{S}_n^0=\emptyset$ and $k=1$}
 	\put(-250,162){Find the best candidate}
 	\put(-220,152){using \eqref{Select}}
 	\put(-255,82){Is the termination criterion}
 	\put(-220,72){fulfilled ?}
 	\put(0,100){$k=k+1$}
 	\put(-250,14){Stop the algorithm and}
 	\put(-220,4){return $\mathcal{S}_n^{k-1}$}
 	\put(-90,83){define}
 	\put(-100,72){$\mathcal{S}_n^k=W_n^k\cup\mathcal{S}_n^{k-1}$}
 	\caption{Flow chart of the NUE algorithm.}
 	\label{NUEFlow}
 \end{figure}
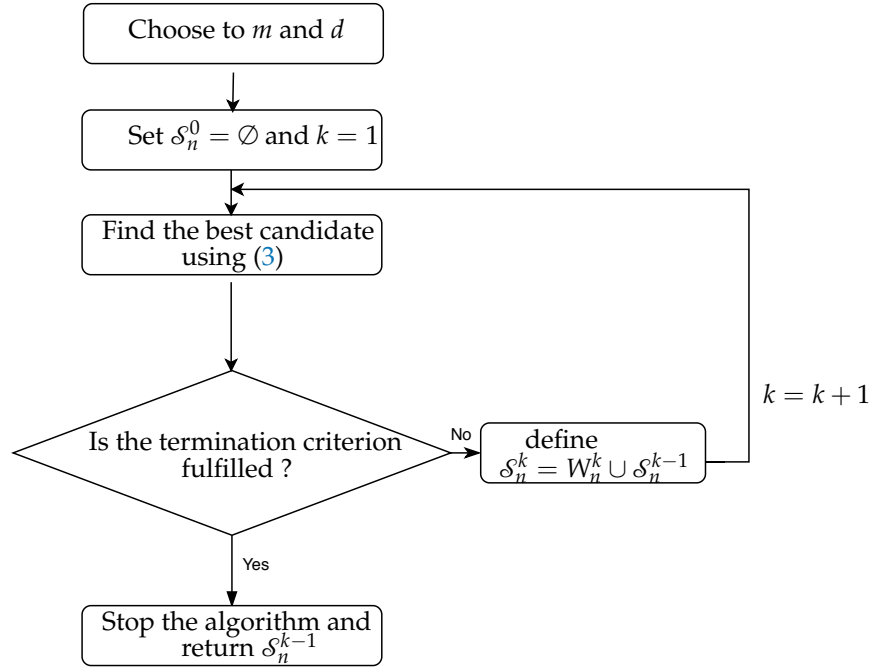

The LA-based strategy follows the same flow chart as the existing NUE algorithm, shown in Figure \ref{NUEFlow}, except that the CMI in \eqref{Select} is substituted by its LA \cite{zhang2018low}. It is suggested in \cite{jia2019detecting,zhang2018low} that using LA of the CMI in \eqref{Select} can increase the accuracy of estimation of the CMI and may outperform the accuracy of the NUE algorithm. The author in \cite{zhang2018low} proposed two LA alternatives to the CMI and concluded based on a simulation study that the LA of the CMI used in this study for the sake of comparison with our proposed NUE algorithm, outperforms another LA of the CMI. The criterion for finding the most informative candidates (i.e. equation \eqref{Select}) in the LA-based NUE algorithm is then given by 

\begin{equation}
W_n^k=\underset{W_n\in \mathcal{C}\setminus \mathcal{S}_n^{k-1}}{\operatorname{argmax}}\left\{ I\left(W_n;Y_n\right)-\frac{2}{|\mathcal{S}_n^{k-1}|}\underset{W_j\in \mathcal{S}_n^{k-1}}{\sum I\left(W_n;W_j\right)}+\frac{2}{|\mathcal{S}_n^{k-1}|}\underset{W_j\in \mathcal{S}_n^{k-1}}{\sum I\left(W_n;W_j|Y_n\right)}\right\},
\label{LA}
\end{equation}

where $|.|$ denotes the cardinality of a set. The mutual information and CMI are estimated using the KSG approach \cite{kraskov2004estimating,montalto2014mute,faes2015estimating} (cf. Appendix \ref{NN}). The LA-based NUE algorithm also uses the bootstrap-based termination criterion. It should be noted that the LA of the CMI (i.e. equation \eqref{LA}) is used to estimate $I^{95}$.

\subsubsection{Akaike Information Criterion-based Non-uniform Embedding Algorithm}
 AIC is used to assess the trade-off between accuracy and complexity of a model. It was adapted to quantify the trade-off between accuracy and complexity of a KDE-based prediction as an alternative to the bootstrap termination criterion in an input variable selection approach in \cite{may2008non,li2015improved}. AIC can also be adapted to act as a termination criterion for stopping the NUE algorithm. Therefore, an AIC-based NUE algorithm could follow the same flow chart as the existing NUE algorithm, shown in Figure \ref{NUEFlow}, except that the the termination criterion will be replaced with the AIC-based termination criterion as is described below.
 
 After selecting the most informative candidate $W_n^k$ by using \eqref{Select}, the target variable $Y_n$ is predicted given $\mathcal{U}_n^k=[W_n^k,\mathcal{S}_n^{k-1}]\in \mathbb{R}^{k}$, by using KDE-based prediction (cf. Appendix \ref{KDE}). Let $y_n=(y_n(1), y_n(1), \dots,y_n(N))$ be N realizations of $Y_n$. The AIC at iteration $k$ is then given as:
 
\begin{equation}
AIC_k=N\log\left(\frac{1}{N}\sum_{i=1}^N(y_n(i)-\widehat{y}_n(i|\mathcal{U}_n^k))^2\right)+2p,
\end{equation}
where the $i^{th}$ realization of $Y_n$ is denoted by $y_n(i)$ and $\widehat{y}_n(i|\mathcal{U}_n)$ is an estimator for the prediction of $y_n(i)$ given $\mathcal{U}_n$. The total number of realization of $Y_n$ is $N$ and $p$ is the measure of complexity and for KDE-based regression, it is given as \cite{may2008non,danafar2014kernel}:
\begin{equation}
p=\sum_{n=1}^N\frac{K_h(\mathcal{u}_n^k(i),\mathcal{u}_n^k(i))}{\sum_{j=1}^NK_h(\mathcal{u}_n^k(i),\mathcal{u}_n^k(j))},
\end{equation}
where $\mathcal{u}_n^k(i)$ is $i^{th}$ realization of $\mathcal{U}_n^k$ (see equation (7) for more details) and $K_h$ is a Gaussian kernel with Mahalonobis distance and Gaussian reference kernel bandwidth (cf. Appendix \ref{KDE}). During the NUE algorithm, if $AIC_k>AIC_{k-1}$ then, $W_n^k$ is included in the embedding vector $\mathcal{S}_n^k$. Otherwise, the algorithm stops and $\mathcal{S}_n^{k-1}$ will be considered as the desired reconstructed past state of the system. 
\section{Proposed Termination Criterion}
In this section, inspired by \cite{may2008non}, we present a new termination criterion. Our proposed criterion is based on non-linear prediction of the target variable, similar to the AIC approach. We modify NN-based regression \cite{altman1992introduction} in order to be able to assess the effect of the selected candidate $W_n^k$ on the accuracy of the prediction of $Y_n$. 

We are interested in non-linear prediction of the random variable $Y_n$ given the random vector $\mathcal{U}_n^k=[W_n^k,\mathcal{S}_n^{k-1}]\in \mathbb{R}^k$. We denote the set of $N$ realizations of $W_n^k$ by $\mathcal{w}_n^k=(w_n^k(1),w_n^k(2)...,w_n^k(N))$ and set of $N$ realizations of $\mathcal{U}_n^k$ be the $N\times k$ matrix

\begin{equation}
\mathcal{u}_n^k=\begin{bmatrix}
w_n^k(1)&w_n^{k-1}(1)&\cdots &w_n^1(1) \\
w_n^k(2)&w_n^{k-1}(2)&\cdots &w_n^1(2) \\
\vdots & \vdots & \ddots & \vdots\\
w_n^k(N)&w_n^{k-1}(N)&\cdots &w_n^1(N)
\end{bmatrix}
.
\end{equation}

The $i^{th}$ row of the matrix $\mathcal{u}_n^k$ is a realization of the random vector  $\mathcal{U}_n^k$. Let $\mathcal{t}(i)$ be the set of indices of the $T$ nearest neighbors of the $i^{th}$ realization of $\mathcal{U}_n^k$. For example, $\mathcal{t}(i)=\{3,7,9\}$ shows that $3^{rd}$, $7^{th}$, and $9^{th}$ rows of $\mathcal{u}_n^k$ are the $T=3$ nearest neighbors of its $i^{th}$ row. The Euclidean distance is used as the distance metric for finding the nearest neighbors in the NN-based prediction. The prediction of the $i^{th}$ realization of $Y_n$ (i.e. $y_n(i)$) given $\mathcal{U}_n^k$ is then calculated as an average of the realizations of $Y_n$ whose indices are specified by the neighbor search in $\mathcal{u}_n^k$. The average of the $y$-values having the same conditioned past is not an optimal estimator. However, it is simple, works well in the cases that we have considered, and has also been used in previous work on non-conditional NN-based prediction. The $\widehat{y_n}(i|\mathcal{U}_n^k)$ is given as: 

\begin{equation}
\widehat{y_n}(i|\mathcal{U}_n^k)\triangleq\frac{1}{T}\sum_{v\in \mathcal{t}(i)}y_n(v).
\label{Predict}
\end{equation}

For example, if $\mathcal{t}(i)=\{3,7,9\}$ then $\widehat{y}(i|\mathcal{U}_n^k)$ is equal to the mean of $\{y_n(3),y_n(7),y_n(9)\}$. The residual $r(i|\mathcal{U}_n^k)$ can be computed as:
\begin{equation}
r(i|\mathcal{U}_n^k)=y_n(i)-\widehat{y_n}(i|\mathcal{U}_n^k).
\end{equation}

In the NUE algorithm, the most informative candidate at iteration $k$, $W_n^k$, will be included in the embedding vector, if it significantly improves the accuracy of the prediction of the target variable $Y_n$ given $\mathcal{U}_n^k$ compared to the prediction accuracy from the iteration $k-1$. The accuracy of the prediction can be calculated as the mean of the squared prediction residual (MSR):

\begin{equation}
\operatorname{MSR}(Y_n\mid\mathcal{U}_n^k)=\frac{1}{N}\sum_{i=1}^{N}r(i|\mathcal{U}_n^k)^2,
\label{MSE}
\end{equation}
where the smaller MSR, the better prediction. 

We first assume that the NUE algorithm contains at least $k=2$ iterations and the termination test is performed from the second iteration. Accordingly, at each iteration $k\ge2$, if $\operatorname{MSR}(Y_n| \mathcal{U}_n^{k-1})-\operatorname{MSR}(Y_n| \mathcal{U}_n^k)>\gamma$, then $W_n^k$ is included in $\mathcal{S}_n^k$ and the algorithm proceeds to search for more candidates at iteration $k+1$. Otherwise, the algorithm ends and $\mathcal{S}_n^{k-1}$ is considered as the desired embedding vector. The non-negative parameter $\gamma$ defines how much the accuracy of the prediction needs to be improved before a variable is selected. Basically, by increasing the non-negative parameter $\gamma$ which we have introduced, our proposed algorithm terminates sooner, and hence less variables are selected. In other words, the parameter $\gamma$ controls the balance between true positives and true negatives, which can be useful, for example, in taking care of the confounder effects like IC. We will show in Section 5.2.2 that, by choosing a proper $\gamma$ value, the number of true negatives significantly increases while the number of true positives does not decrease significantly in data in which the IC may cause spurious detection of directed dependencies.

\section{Proposed Non-uniform Embedding Algorithm} 
Our proposed NUE algorithm (referred to as MSR-based) uses a weighted combination of the CMI and MSR for selecting the most informative candidate and our proposed termination criterion for ending the algorithm. The details of the proposed NUE algorithm are as follows: 
\begin{enumerate}[leftmargin=*,labelsep=4.9mm]
\item	Choose $\gamma$, $\lambda$, embedding delay $d$ and embedding dimension $m$ and construct the candidate set $\mathcal{C}=[X_{n-m},...,X_{n-md},Y_{n-m},..Y_{n-md},\textbf{Z}_{n-m},...,\textbf{Z}_{n-md}]$.  
\item	Initialize by setting $\mathcal{S}_n^0=\emptyset$,
\item	At first iteration $k=1$, find the first most relevant candidate $W_n ^1$ by using a weighted combination of MSR and mutual information as:
 
\begin{equation}
W_n^1=\underset{W_n\in \mathcal{C}}{\operatorname{argmax}}\left[(1-\lambda)\operatorname{I}\left(Y_n;W_n\right) -\lambda \operatorname{MSR}(Y_n\mid W_n\right)],
\label{SFC}
\end{equation}

where $0\leq\lambda\leq1$ is the weight. Then set $\mathcal{S}_n^1=[W_n^1]$.
\item At each iteration $k\geq2$, run a search procedure to select the candidate which leads to the highest amount of new information about target variable $Y_n$ and the best prediction of $Y_n$ given the random vector  $\mathcal{U}_n^k=[W_n,\mathcal{S}_n^{k-1}]$. It can be formalized by: 
 
\begin{equation}
W_n^k=\underset{W_n\in \mathcal{C}\setminus \mathcal{S}_n^{k-1}}{\operatorname{argmax}}\left[(1-\lambda)\operatorname{I}\left(Y_n;W_n\mid{\mathcal{S}_n^{k-1}}\right) -\lambda \operatorname{MSR}\left(Y_n\mid \mathcal{U}_n^k\right)\right],
\label{SRC}
\end{equation}

where $\mathcal{S}_n^{k-1}=\bigcup\limits_{i=0}^{k-1}W_n^i$ denotes the set of selected candidates up till iteration $k-1$ and $\mathcal{C}\setminus \mathcal{S}_n^{k-1}$ refers to all elements of $\mathcal{C}$ except the elements of $\mathcal{S}_n^{k-1}$. Similar to the existing NUE algorithms, mutual information and CMI are estimated using the KSG approach \cite{kraskov2004estimating,montalto2014mute,faes2015estimating} (cf. Appendix \ref{NN}).
\item  Include the candidate $W_n^k$ in the embedding vector $\mathcal{S}_n^k$ if $\operatorname{MSR}(Y_n| \mathcal{U}_n^{k-1})-\operatorname{MSR}(Y_n| \mathcal{U}_n^k)>\gamma$ and continue the algorithm to find more candidate. Otherwise, terminate the algorithm and return  $\mathcal{S}_n^{k-1}$ as the desired embedding vector.		
\end{enumerate}

\begin{figure}[t]
	\centering
	\includegraphics[width=10cm,trim={0 0.01cm 0 0.01cm},clip]{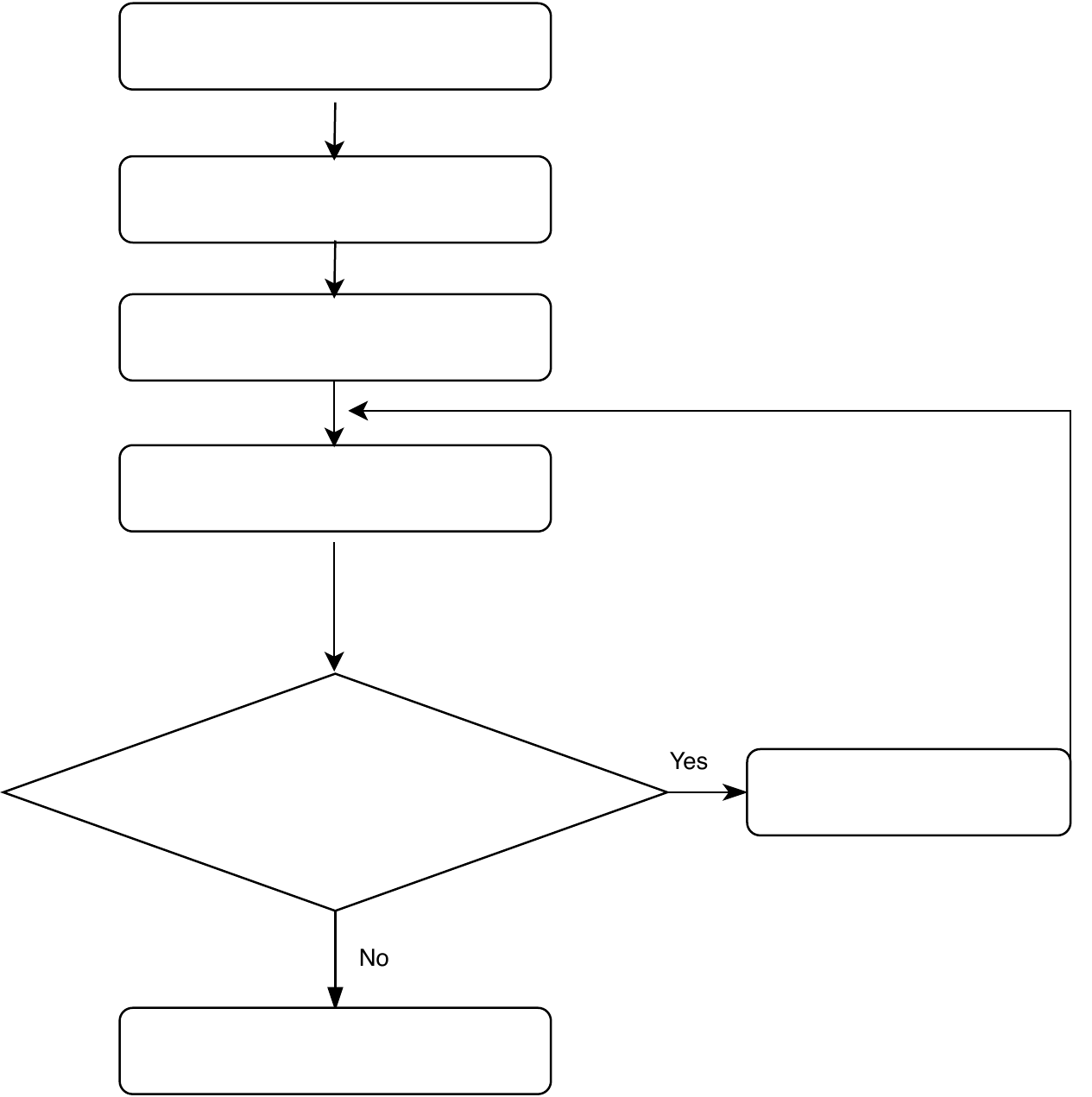}
	\put(-240,270){Choose $m$, $d$, $\gamma$ and $\lambda$}
	\put(-240,230){Set $\mathcal{S}_n^0=\emptyset$ and $k=1$}
	\put(-240,199){Find $W_n^1$ using \eqref{SFC}}
	\put(-250,160){Find the best candidate}
	\put(-220,150){using \eqref{SRC}}
    \put(-275,76){$\operatorname{MSR}(Y_n| \mathcal{U}_n^k)-\operatorname{MSR}(Y_n| \mathcal{U}_n^{k-1})<\gamma$}
    \put(-70,83){define}
    \put(-80,72){$\mathcal{S}_n^k=W_n^k\cup\mathcal{S}_n^{k-1}$}
    \put(0,100){$k=k+1$}
    \put(-250,14){Stop the algorithm and}
    \put(-220,4){return $\mathcal{S}_n^{k-1}$}
	\caption{The flow chart of our proposed NUE algorithm.}
	\label{NNUEFlow}
\end{figure}
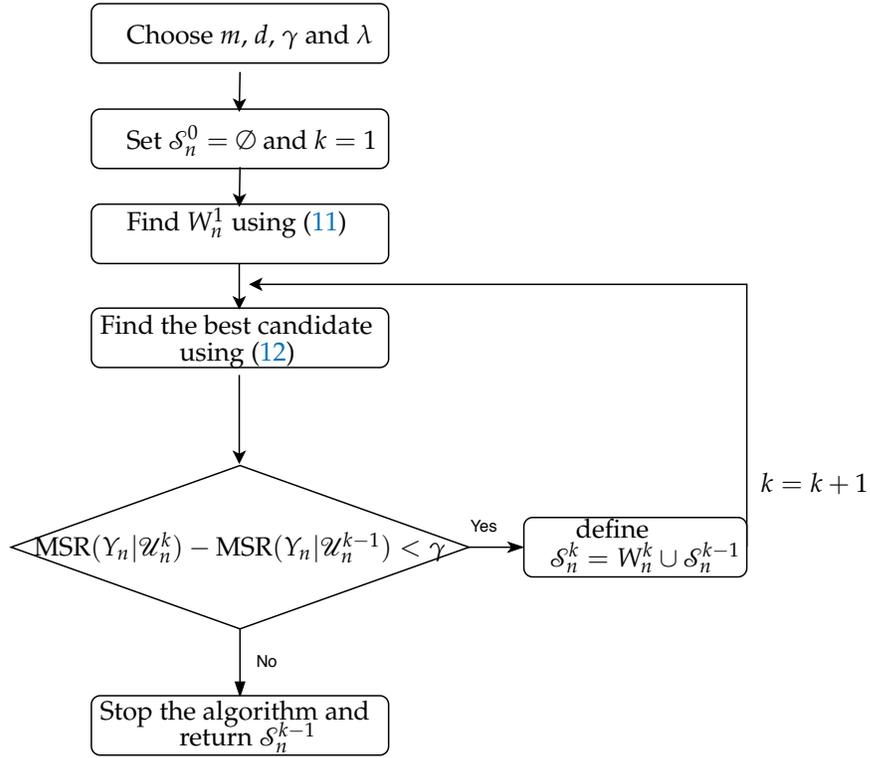

The flow chart of the proposed algorithm is shown in Figure \ref{NNUEFlow}. CTE is then estimated by replacing $[X_n^-,Y_n^-,\textbf{Z}_n^-]$ and $[Y_n^-,\textbf{Z}_n^-]$ with $\mathcal{S}_n^{k-1}$ and $\mathcal{S}_n^{k-1}$ excluding the past of $X_n$, respectively. The CTE is finally estimated using the KSG approach \cite{kraskov2004estimating,montalto2014mute,faes2015estimating} (cf. Appendix \ref{NNPTE}).
 
\section{Simulation Study}
\label{S4}
In this section, we use simulated data in order to compare the performance of our proposed NUE algorithm with the existing algorithms described in Section 2.2. We investigate the effect of the data length, strength of directed dependency and instantaneous coupling effect on the NUE algorithms. The execution time of the NUE algorithms are also investigated. 
The main reason for using simulated data is to be able to obtain well-defined ground truth. Therefore, it is possible to compare the NUE algorithms by computing their accuracies. The termination criterion of the NUE algorithms is also utilized for testing the significance of the estimated CTE in the simulation study: if the embedding vector $\mathcal{S_n}$ of the target variable $Y_n$ does not include any lagged component of the node $\mathcal{X}$, then CTE from $\mathcal{X}$ to $\mathcal{Y}$ is zero and, otherwise its CTE is positive. The results are used to calculate true positive (TP), i.e number of truly detected directed coupled nodes, true negative (TN), false positive (FP), and false negative (FN). The accuracy (ACC), true positive rate (TPR), and true negative rate (TPR) of the NUE algorithms are then defined as:

\begin{equation}
\begin{array}{l}
\operatorname{ACC}=100\times\frac{TP+TN}{TP+TN+FP+FN}\vspace{0.1cm}\\
\operatorname{TNR}=100\times\frac{TN}{TN+FP}\vspace{0.1cm}\\
\operatorname{TPR}=100\times\frac{TP}{TP+FN}.
\end{array}
\label{ACC}
\end{equation}

The TPR shows the ability of NUE algorithms to include the candidates in the embedding vector related to correctly coupled nodes, and TNR represents the ability to exclude the candidates related to uncoupled nodes. The ACC, TPR and TNR are computed as an average over $100$ generated realizations because the simulated data depends on the random initial condition. The embedding delay $m$ and dimension $d$ are chosen as $1$ and $5$ samples, respectively. For estimation of the CMI and MSR,  $T=10$ nearest neighbors are considered. 
\subsection{Henon Map Model} \label{SD}

The Henon map model has been frequently utilized in the literature to generate multivariate data with a controlled amount of directed interaction \cite{montalto2014mute,xiong2017entropy,zhang2018low}. A 5 nodes Henon map can be defined as \cite{montalto2014mute,xiong2017entropy,zhang2018low}:

\begin{equation}
\begin{array}{ll}
Y_{l,n} & =1.4-Y^2_{l,n-1}+0.3Y_{l,n-2},\;\;\;\mbox{for $l=1,5$} \vspace{0.1cm}\\
Y_{l,n} & =  1.4-\left[0.5Q(Y_{l-1,n-1}+Y_{l+1,n-1})+(1-Q)Y_{l,n-1}\right]^2+0.3Y_{l,n-2}\vspace{0.1cm}
\hspace{.1cm},\;\;\;\mbox{for $l=2,3,4$},
\label{Hen}
\end{array}
\end{equation}

where $Q$ is the coupling strength and it varies between 0.2 to 0.8 in this study; it is guaranteed that the complete synchronization between any pair nodes is avoided \cite{faes2014lag}. The first and last nodes ($Y_1$ and $Y_5$) depend only on their own past (first row of \eqref{Hen}) and therefore they do not depend on other nodes. On the other hand, nodes $l=2,3,4$ depend on the past of nodes $Y_{l-1}$ and $Y_{l+1}$. Consequently, there are non-linear directed dependencies with strength $Q$ from nodes $Y_{l-1}$ and $Y_{l+1}$ to node $Y_{l}$ for $l=2,3,4$ (second row of \eqref{Hen}). The aforementioned connectivity is considered as the ground truth when comparing the performance of the NUE algorithms.
\subsubsection{Data Length Effect}
Henon map data sequences were generated at a fixed normal strength $Q=0.6$ and different lengths, $N=2^h, h=5,6,...,10$, in order to evaluate the effect of the data length on the performance of the NUE algorithms. The proposed NUE algorithm were used with five different weights, $\lambda=0,0.25,0.5,0.75,1$, to demonstrate the effect of the weight. According to the fact that in this simulation there is no unobserved confounder effect like IC, we set the parameter $\gamma=0$. 
\begin{figure*}[t]
	\centering
	\begin{subfigure}[t]{0.3\linewidth}
		\includegraphics[width=5cm,height=5cm]{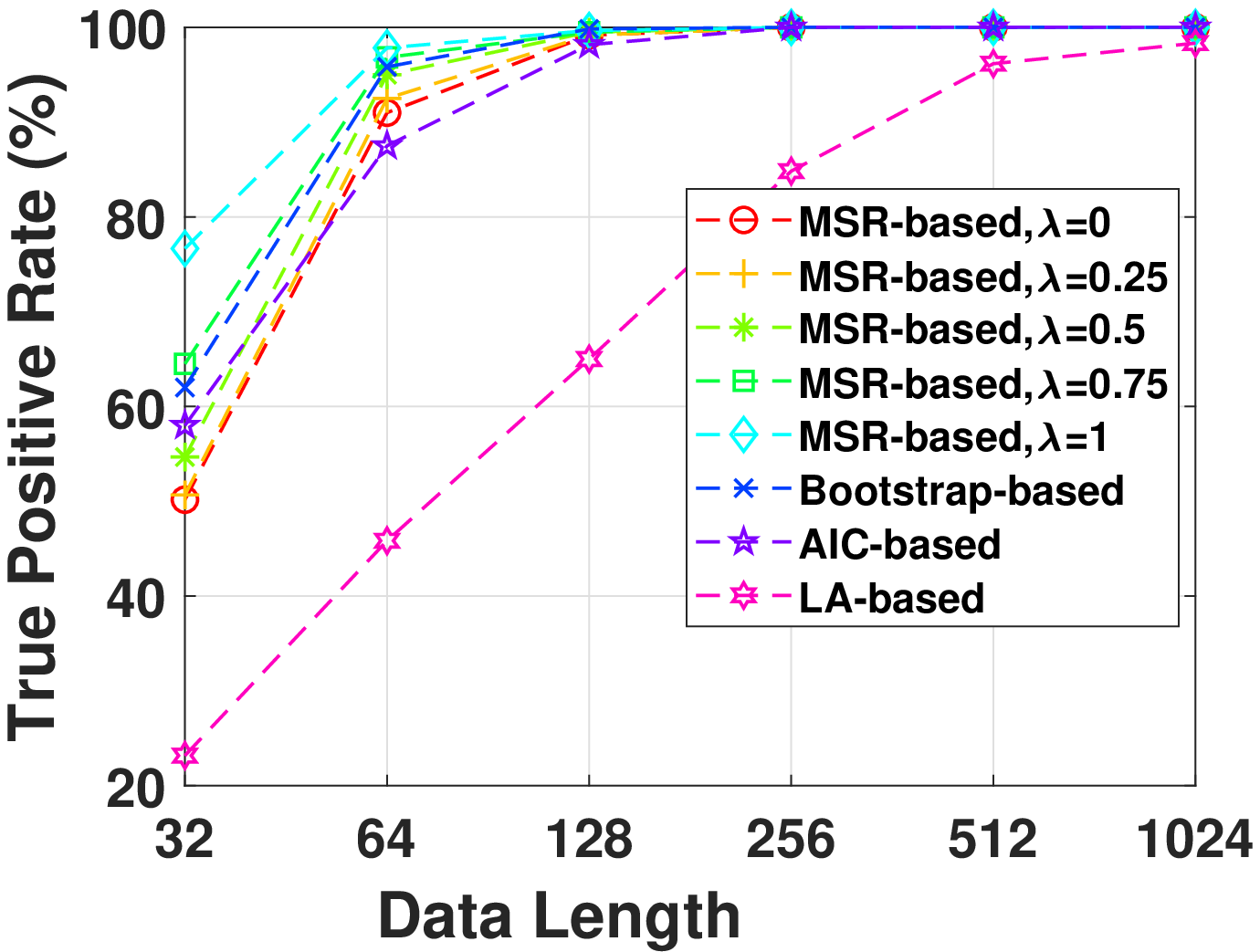}
		\caption{}
		\label{TPR_HM}
	\end{subfigure}
	\begin{subfigure}[t]{0.3\linewidth}
		\includegraphics[width=5cm,height=5cm]{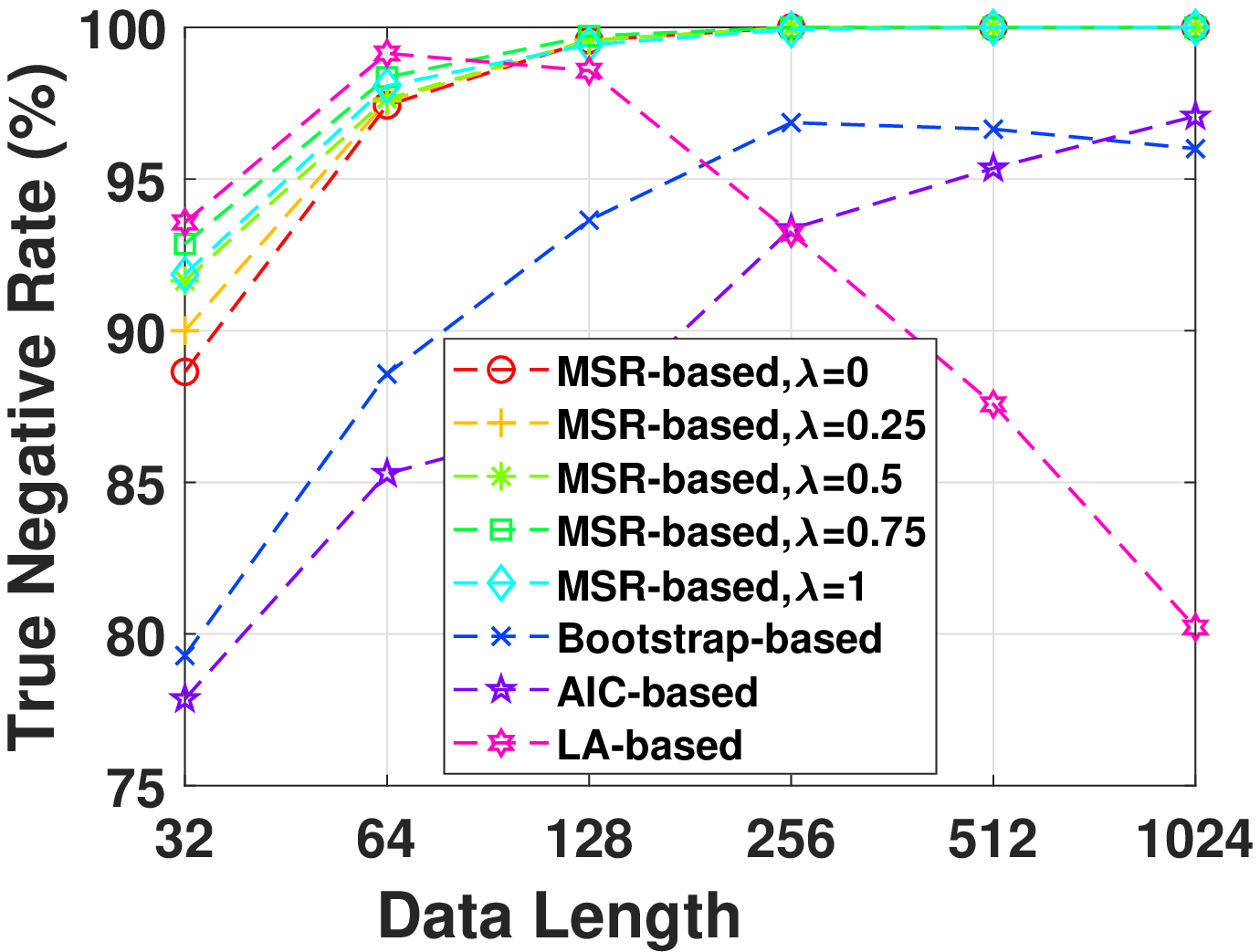}
		\caption{}
		\label{TNR_HM}
	\end{subfigure}
	\begin{subfigure}[t]{0.3\linewidth}
		\includegraphics[width=5cm,height=5cm]{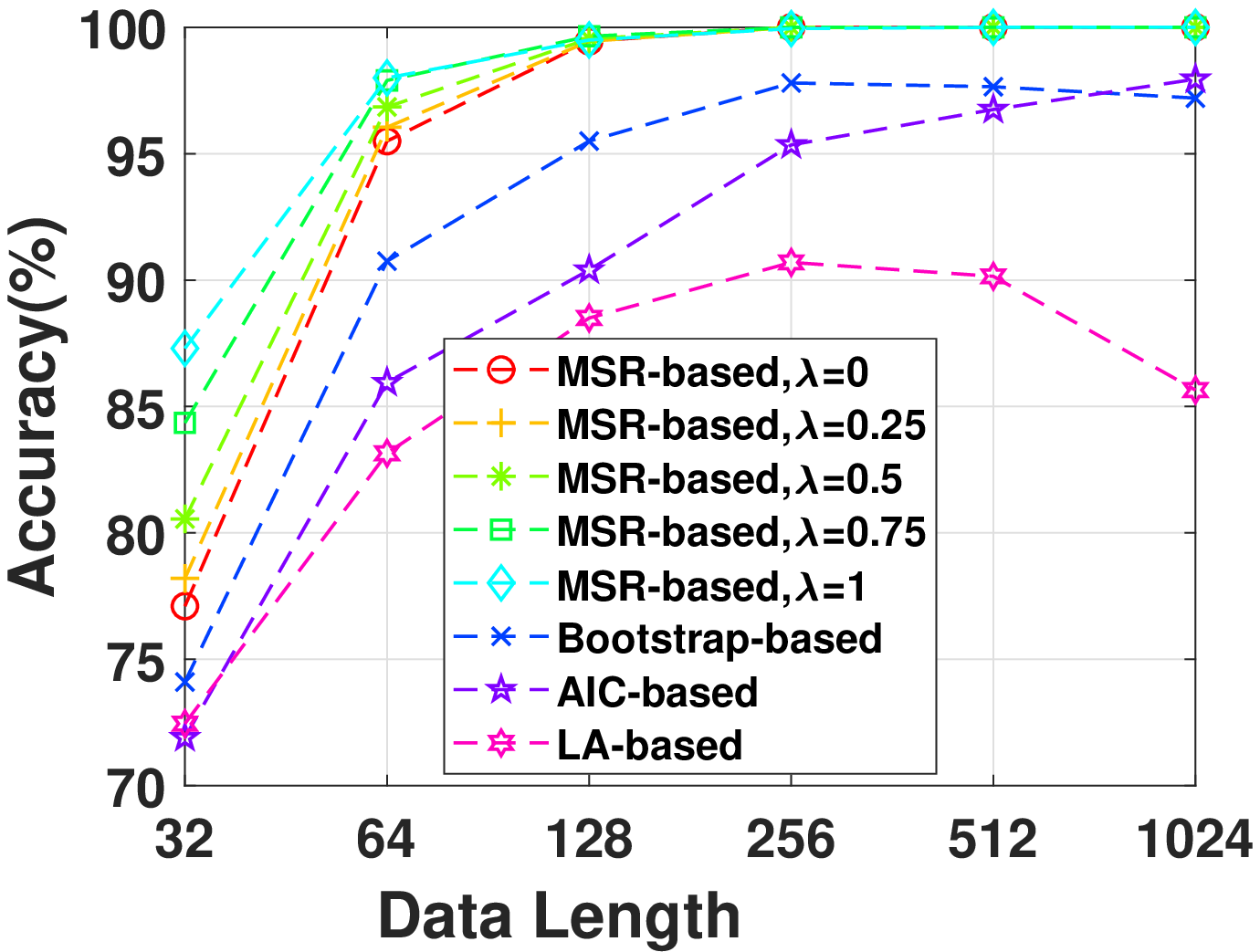}
		\caption{}
		\label{ACC_HM}
	\end{subfigure}
	
	\caption{(a) true positive rates, (b) true negative rates and (c) accuracies of MSR-based, bootstrap-based, AIC-based, and LA-based NUE algorithms for the Henon map model at moderate fixed coupling strength $Q=0.6$ and data length ranging from 32 to 1024. The results are shown as an average over 100 realizations.}
	\label{HM}
\end{figure*}
Figure \ref{HM} shows TPRs, TNRs and accuracies of the MSR-based NUE algorithm with five different $\lambda$'s. Also shown in the figure, are the performances of the existing NUE algorithms. As Figure \ref{ACC_HM} demonstrates, the accuracy of our proposed NUE algorithm (for any $\lambda$) increases as the data length increases up to 256 samples where the accuracy is nearly 100\%. The proposed algorithm with  higher $\lambda$ attains better performance at data length under 128 samples. Figures \ref{TPR_HM} and \ref{TNR_HM} show that the improvement of the accuracy by changing $\lambda$ is mostly due to the better TPRs. As we can see in Figure \ref{TNR_HM}, TNRs of bootstrap-based and LA-based algorithms decreases for data lengths greater than 256 and 64, respectively. The accuracy, TPR and TNR of the AIC-based algorithm increases by increasing the data length. Overall, the proposed algorithm with $\lambda=1$ attains the greatest accuracy and the LA-based algorithm has the worst accuracy for all data lengths. 
\subsubsection{Coupling Strength Effect}
\begin{figure*}[t]
	\centering
	\begin{subfigure}[t]{0.3\linewidth}
		\includegraphics[width=5cm,height=5cm]{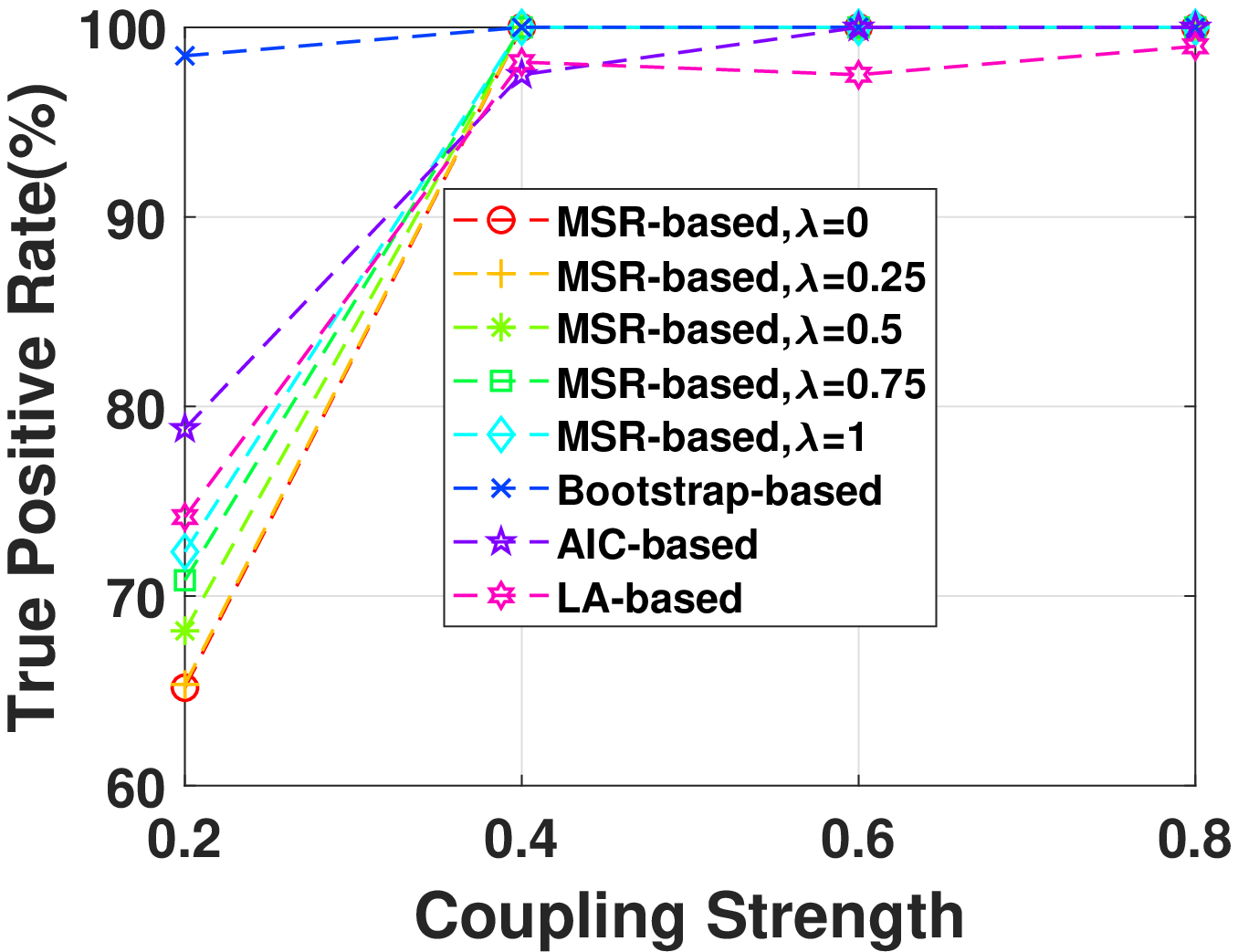}
		\caption{}
		\label{TPR_HM_S}
	\end{subfigure}
	\begin{subfigure}[t]{0.3\linewidth}
		\includegraphics[width=5cm,height=5cm]{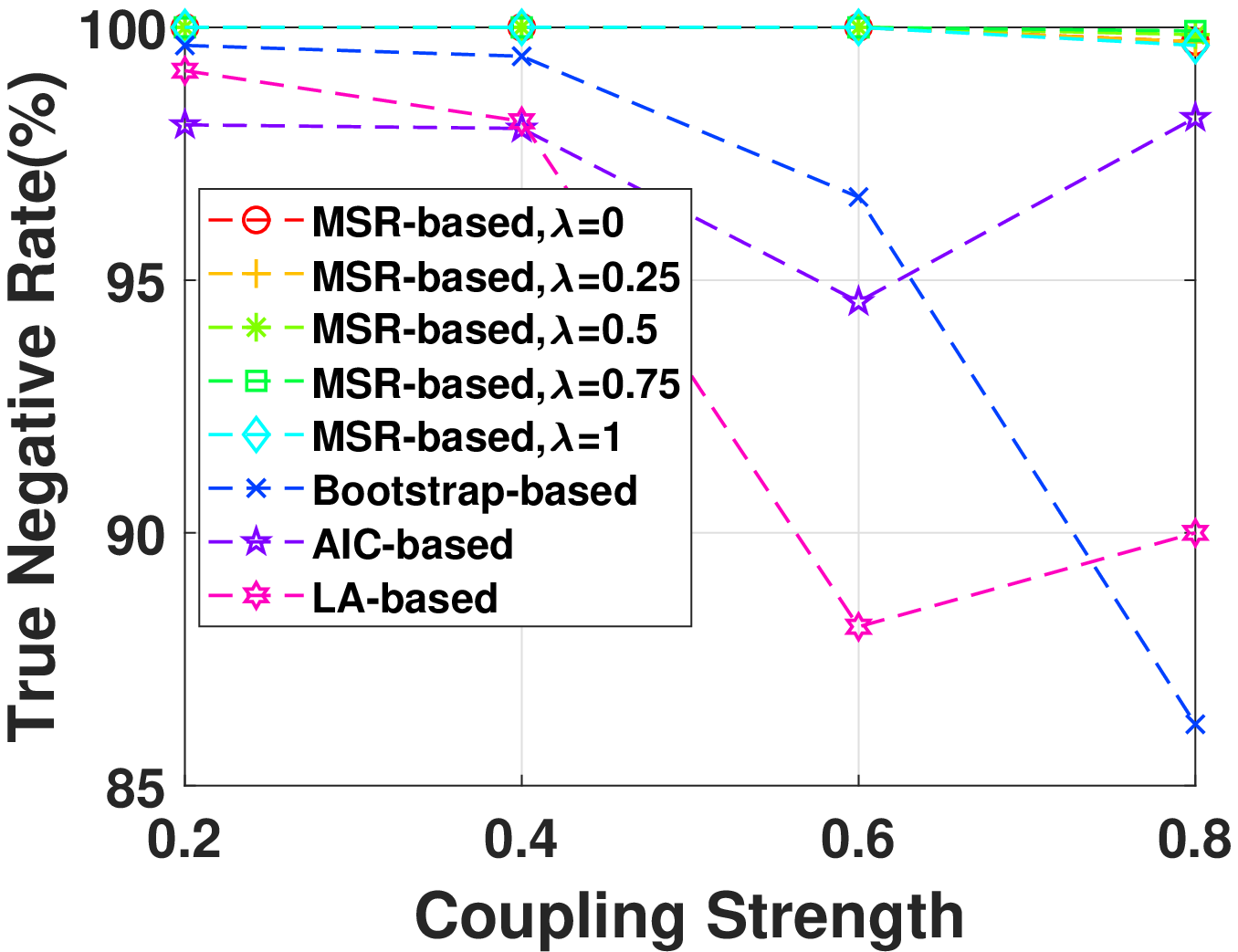}
		\caption{}
		\label{TNR_HM_S}
	\end{subfigure}
	\begin{subfigure}[t]{0.3\linewidth}
		\includegraphics[width=5cm,height=5cm]{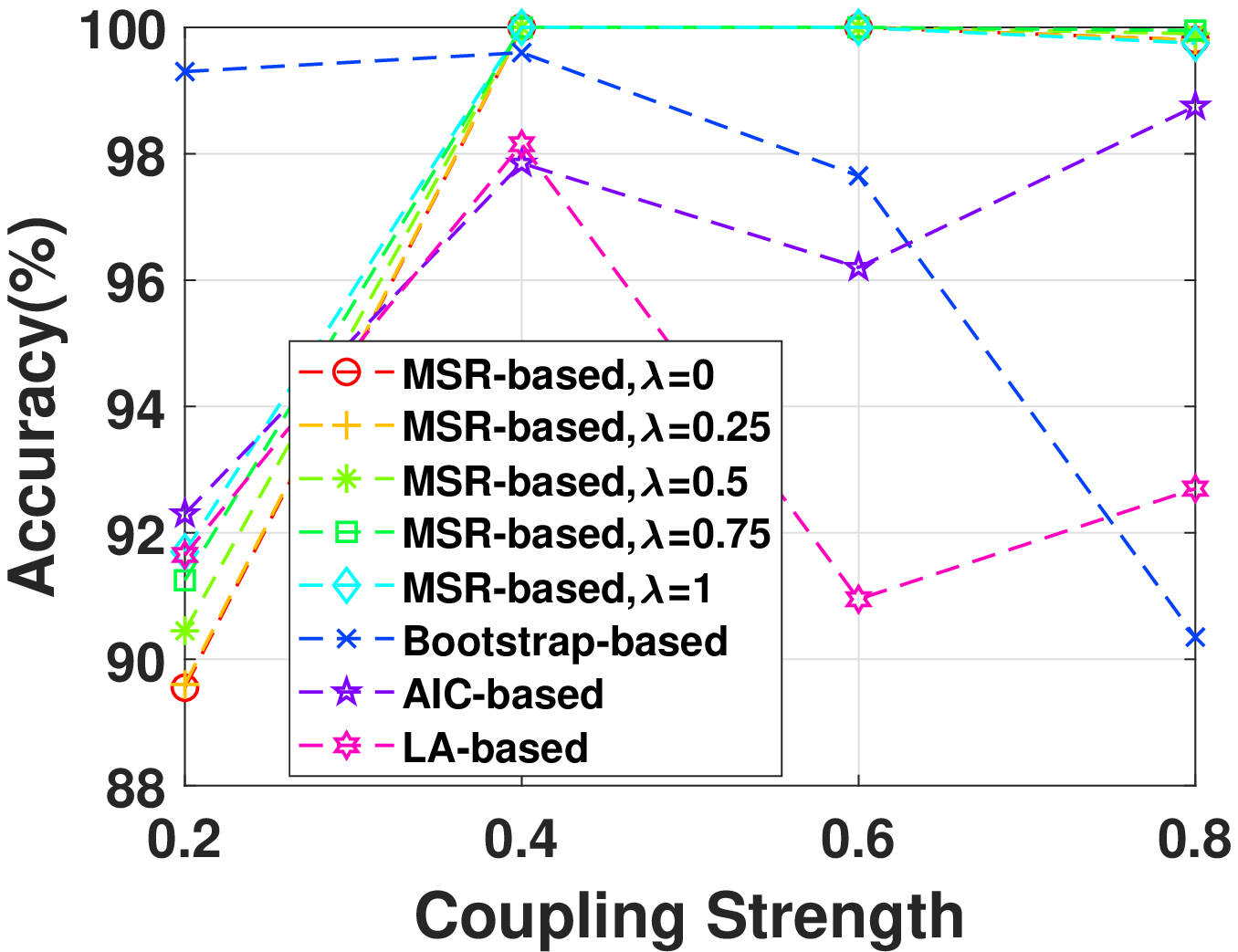}
		\caption{}
		\label{ACC_HM_S}
	\end{subfigure}
	
	\caption{(a) true positive rates, (b) true negative rates and (c) accuracies of MSR-based, bootstrap-based , AIC-based, and LA-based NUE algorithms for the Henon map model at fixed data length $N=512$ and coupling strength ranging from 0.2 to 0.8. The results are shown as an average over 100 realizations.}
	\label{HM_Strength}
\end{figure*}

The Henon map model at 512 data length was generated with different coupling strengths ranging from 0.2 to 0.8 in step of 0.2 in order to evaluate the NUE algorithms as a function of the strength of the directed dependencies. As Figure \ref{TNR_HM_S} shows TNRs of the MSR-based algorithm (for any $\lambda$) is almost 100 \% while the TNRs of the existing NUE algorithms tend to decrease as the strength of the directed dependency increase, which also causes a decrease in the accuracy. TPRs of the NUE algorithms are nearly equal except that at very low coupling strength the bootstrap-based algorithm has higher TPR. Changing $\lambda$ at $Q=0.2$ leads to slightly better TPR and accuracy.  Overall, our proposed MSR-based algorithm has better accuracy compared to that of the existing NUE algorithms, except for $Q=0.2$ where bootstrap-based algorithms yields better performance.

\subsubsection{Execution Time}
In this section the execution time of the proposed MSR-based algorithm with $\lambda=1$ and $\lambda=0$ (at fixed $\gamma=0$) is compared with that of the existing NUE algorithms. The Henon Map data at length 512 samples and coupling strength $Q=0.6$ was generated and execution time of the NUE algorithms are reported as an average over 100 realizations. The execution time was calculated in a single block-wise code where each NUE algorithms has a block. The function \textit{tic} of MATLAB was set before each block and the function \textit{toc} was used to calculate the execution time of the blocks related to the NUE algorithms. The code was run by a Intel(R) core(TM) i7-7600 CPU @ 2.10 GHz. We use the ITS toolbox (available at http://www.lucafaes.net/its.html) for implementation of bootstrap-based NUE algorithm. The ITS toolbox was also modified for implementation of the LA-based algorithm by using a MATLAB code provided in \cite{zhang2018low}. We also modified ITS toolbox in order to implement the AIC-based and MSR-based NUE algorithms. The results are reported in Table \ref{Tab1}. In addition to execution time, the total number of iterations $k$ that the algorithms were performed before they terminated, are reported.  

\begin{table}[H]
	\caption{The execution time and and total iterations before termination of the proposed MSR-based with $\lambda=0,1$ (at fixed $\gamma=0$) as well as existing NUE algorithms for the Henon map data at data length 512. The results are reported as an average over 100 realizations.}
	\label{Tab1}
	\centering
	\begin{tabular}{cccccc}
		\toprule
		
		\textbf{NUE algorithm}  &\textbf{Bootstrap-based} & \textbf{LA-based} & \textbf{AIC-based}&\textbf{MSR-based, $\lambda=1$}&\textbf{MSR-based, $\lambda=0$}\\ 
		\midrule
		\textbf{Execution Time (second)}	&40.59&117.23&11.29&2.26&5.34\\
		\textbf{Total Number of Iterations}	&19.18&16.94&24.08&16.19&16.64\\
		\bottomrule
	\end{tabular}
\end{table}

As Table \ref{Tab1} indicates, the execution time of MSR-based with the known parameters $\lambda=1$ and $\lambda=0$, and AIC-based NUE algorithms are significantly less than that of the bootstrap-based and LA-based ones. However, the total number of iterations of the AIC-based algorithm before termination is on average higher in comparison with that of the MSR-based algorithm. The higher total number of iterations of the AIC-based algorithm increases its execution time. It is important to note that the execution time of the MSR-based with $\lambda=1$ is less than that of with $\lambda=0$. Overall, our proposed MSR-based NUE algorithm with $\lambda=1$ and $\gamma=0$ attains the best and the LA-based has the worst execution time.

\subsection{Autoregressive Model}

AR models have been widely used to generate multivariate data with controlled directed dependencies among them \cite{montalto2014mute,xiong2017entropy,zhang2018low}. The considered non-linear AR model is given as:  
\begin{equation}
\begin{array}{l}
Y_{1,n}=0.95\sqrt{2}Y_{1,n-1}-0.9125Y_{1,n-2}+\varepsilon_1\vspace{0.1cm}\\
Y_{2,n}=0.5Y_{1,n-2}^2+\varepsilon_2\vspace{0.1cm}\\
Y_{3,n}=-0.4Y_{1,n-3}+0.4Y_{2,n-1}+\varepsilon_3\vspace{0.1cm}\\
Y_{4,n}=-0.5Y_{1,n-1}^2+0.25\sqrt{2}Y_{4,n-1}+\varepsilon_4\vspace{0.1cm}\\
Y_{5,n}=-0.25\sqrt{2}Y_{4,n-1}+0.25\sqrt{2}Y_{5,n-2}+\varepsilon_5,
\label{AR}
\end{array}
\end{equation}
where $\varepsilon_{1},..., \varepsilon_{5} $ are mutually independent zero mean and unit variance white Gaussian noise processes. In accordance with \eqref{AR}, node 1 only depends on its own past and therefore there is no directed dependency from other nodes to node 1 (first row of \eqref{AR}). On the other hand, nodes 2, 3 and 4 depend on the past of node 1 and therefore there are non-linear directed dependencies from node 1 and to nodes 2 and 4 (second and fourth rows of \eqref{AR}) and linear directed dependencies from node 1 to node 3 (third row of \eqref{AR}). There are also linear directed dependencies from nodes 2 and 4 to nodes 3 and 5, respectively (third and fifth rows of \eqref{AR}). These dependencies describe the ground truth couplings when comparing TPR, TNR, and ACC of the NUE algorithms.
\subsubsection{Data Length Effect}
\begin{figure*}[t]
	\centering
	\begin{subfigure}[t]{0.3\linewidth}
		\includegraphics[width=5cm,height=5cm]{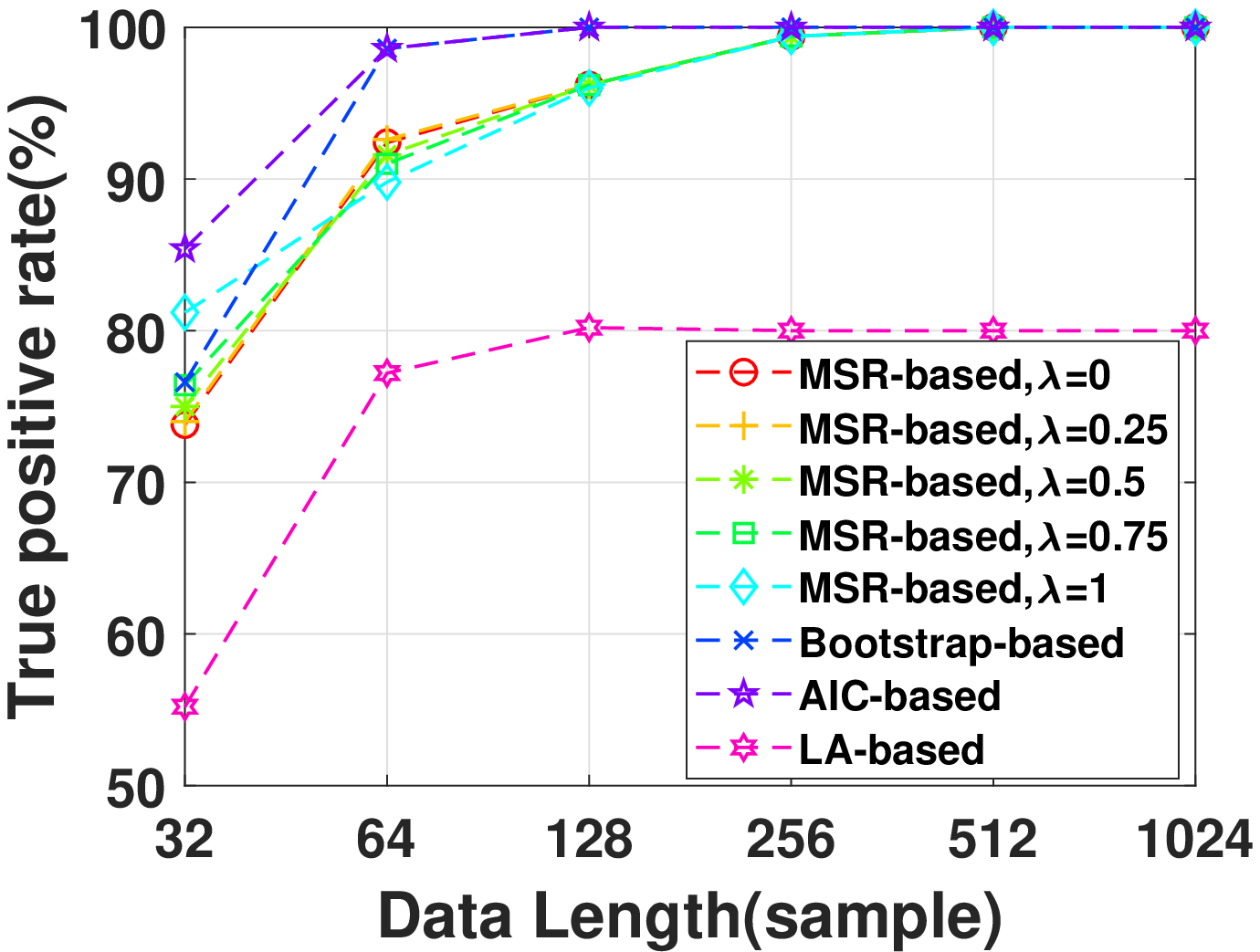}
		\caption{}
		\label{TPR_AR}
	\end{subfigure}
	\begin{subfigure}[t]{0.3\linewidth}
		\includegraphics[width=5cm,height=5cm]{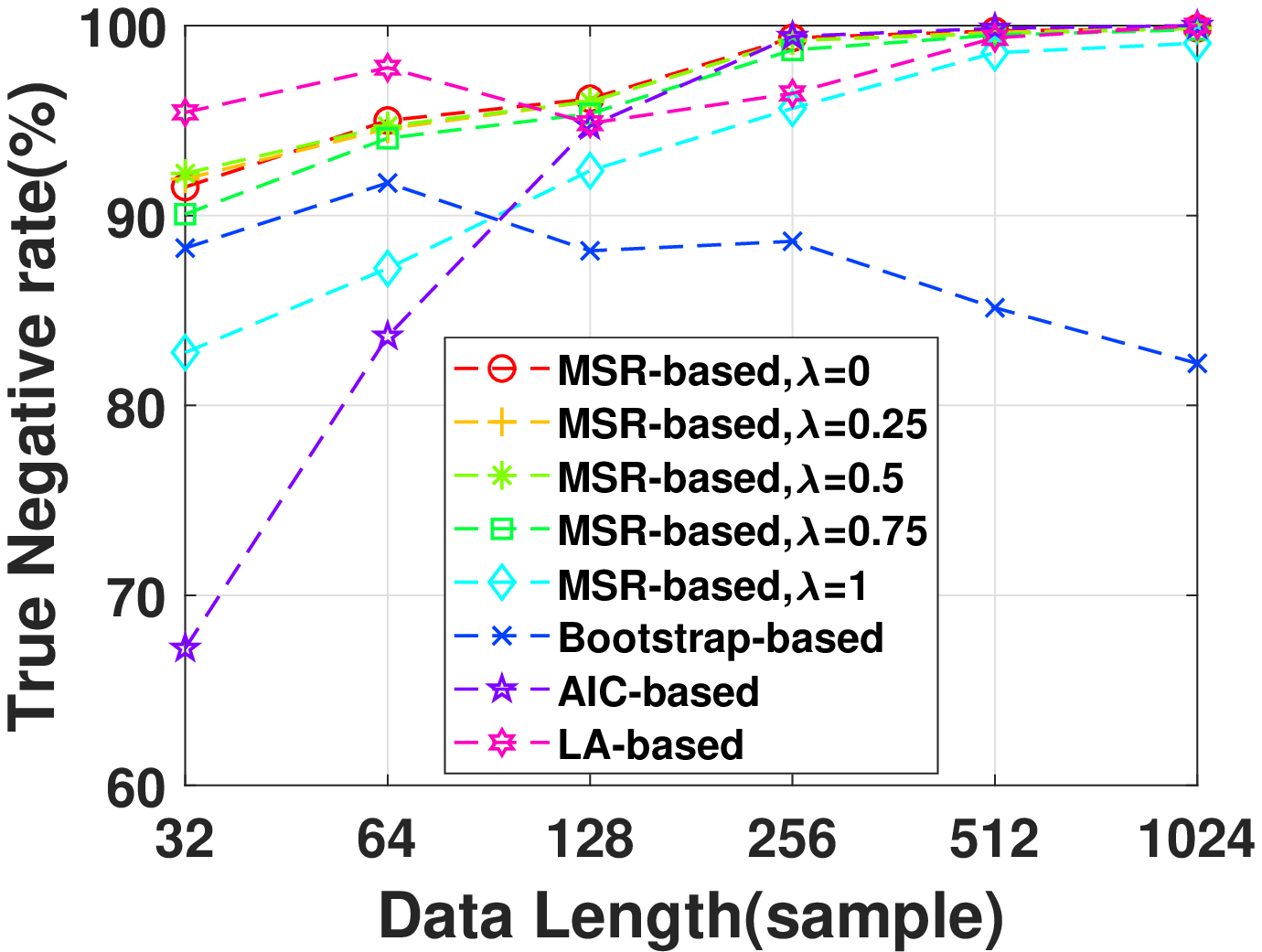}
		\caption{}
		\label{TNR_AR}
	\end{subfigure}
	\begin{subfigure}[t]{0.3\linewidth}
		\includegraphics[width=5cm,height=5cm]{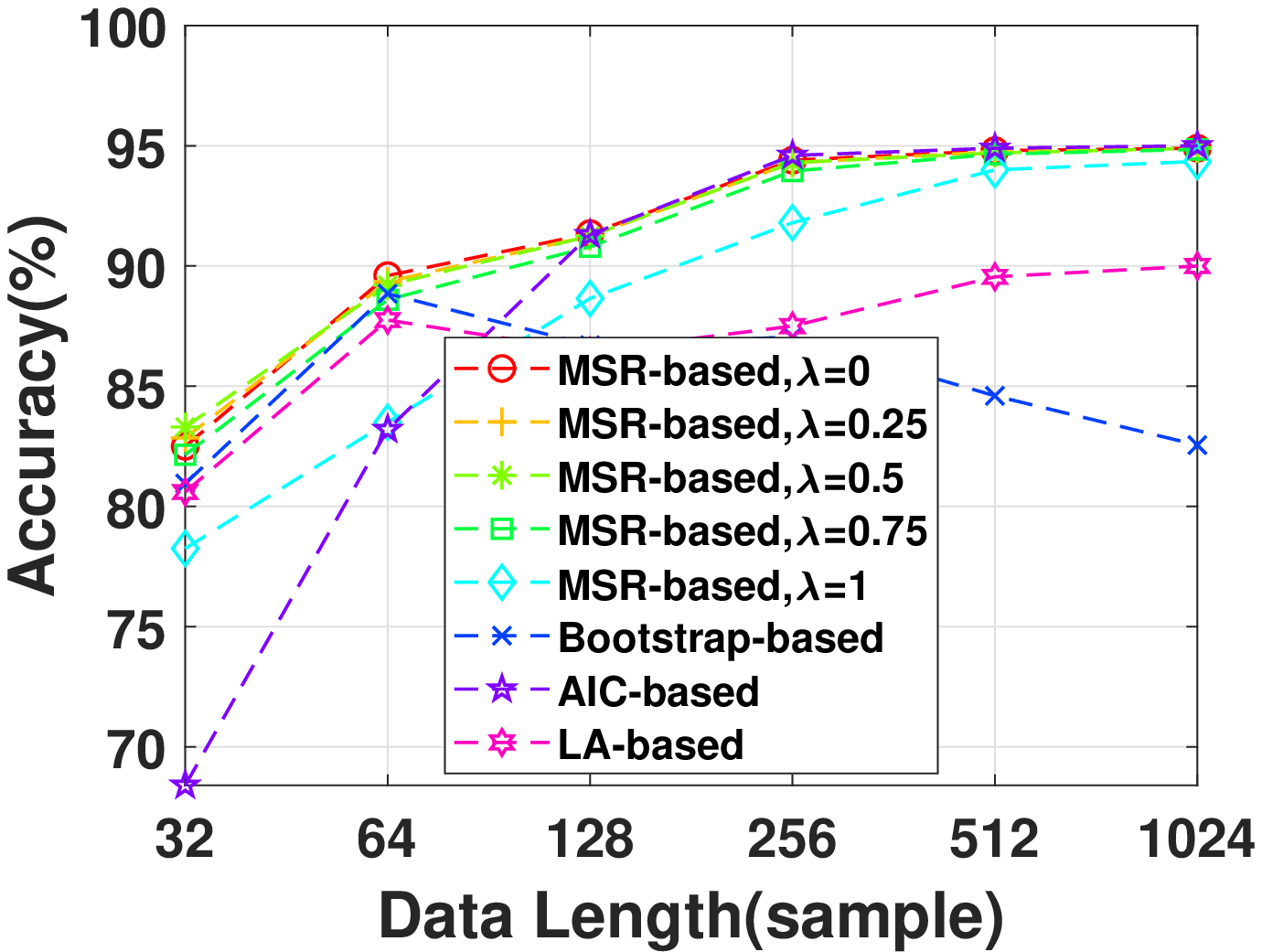}
		\caption{}
		\label{ACC_AR}
	\end{subfigure}
	
	\caption{(a) true positive rates, (b) true negative rates and (c) accuracies of MSR-based, bootstrap-based, AIC-based, and LA-based NUE algorithms for the AR at data length ranging from 32 to 1024. The results are shown as an average over 100 realizations.}
	\label{fig: AR}
\end{figure*}

Non-linear AR data series were first generated for 100 realizations at different lengths, $N=2^h, h=5,6,...,10$, in order to evaluate the effect of data length on the performance of the NUE algorithms using AR data. We set the parameter $\gamma=0$ since in this simulation there is no IC effect. Figure \ref{fig: AR} shows TPRs, TNRs and accuracies of the NUE algorithms for the AR model as a function of data lengths. As figure \ref{TPR_AR} illustrates, the LA-based NUE algorithm has significantly lower TPR compared to that of the other algorithms. It is also noteworthy that the TNR of the bootstrap-based algorithm tends to decrease as the data length increases. The MSR-based algorithm, for all $\lambda$ except $\lambda=1$, presents higher accuracy than that of the bootstrap-based and LA-based algorithms at all data lengths and higher accuracy than that of the AIC-based algorithms at data length smaller than 128.
\subsubsection{Instantaneous Coupling Effect}
IC can happen due to sharing information at same lag. In other words, it can occur due to fast sharing information \cite{faes2013compensated}. For example, in neuro-physiological time series like EEG, the recorded electrical activity at each electrode located at the scalp, is considered to be a mixture of the source generators because the sources pass through the volume conductor \cite{ruiz2019computational}. The volume conduction can be considered as the zero lag coupling which may lead to detection of false directed dependency by the NUE algorithms.

Let us consider the AR model defined in (15) at length $N$ as the sources, which are instantly mixed to simulate the effect of IC. The considered mixing matrix is given as
\begin{equation}
A=\begin{bmatrix}
(1-\alpha)&\alpha&\alpha&\alpha&\alpha \\
\alpha&(1-\alpha)&\alpha&\alpha&\alpha\\
\alpha&\alpha&(1-\alpha)&\alpha&\alpha\\
\alpha&\alpha&\alpha&(1-\alpha)&\alpha\\
\alpha&\alpha&\alpha&\alpha&(1-\alpha)
\end{bmatrix}
.
\end{equation}
where $\alpha$ varies between 0.1 and 0.3 in step of 0.1 in this paper. The greater $\alpha$, the greater IC between the sources. Let $Y=[Y_1,Y_2,\dotsc,Y_5]^T$ be $N \times 5$ matrix which includes all sequences  (they are considered to simulate sources in the brain) generated by the AR model \eqref{AR}. The mixed matrix (it is considered to simulate the EEG signals recorded at the scalp level which is the mixture of all sources) is then defined as the matrix product between $Y$ and $A$, that is 
\begin{equation}
Y^{mixed}=YA.
\label{MIxed}
\end{equation} 

Each column of $A$ defines how the sources $Y_1,\dotsc Y_5$ are mixed. As expected, for the $n^{th}$ mixed data sequence $Y^{mixed}_n$, the most important term is $Y_n$. This is more clear by looking at the main diagonal of the $A$.

\begin{figure*}[t]
 	\centering
 	\begin{subfigure}[t]{0.3\linewidth}
 		\includegraphics[width=5cm,height=5cm]{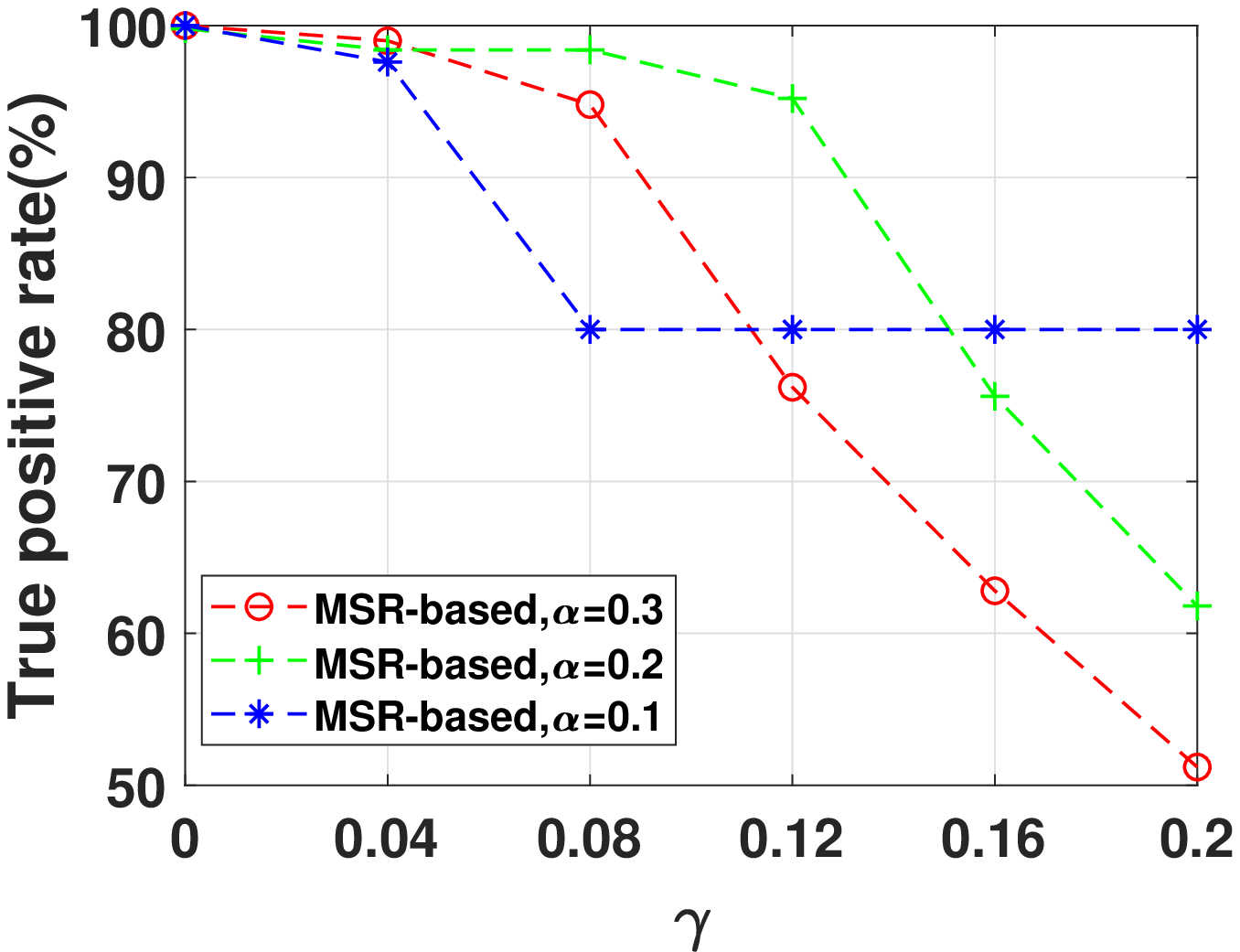}
 		\caption{}
 		\label{Gam_Efe_TPR}
 	\end{subfigure}
 	\begin{subfigure}[t]{0.3\linewidth}
 		\includegraphics[width=5cm,height=5cm]{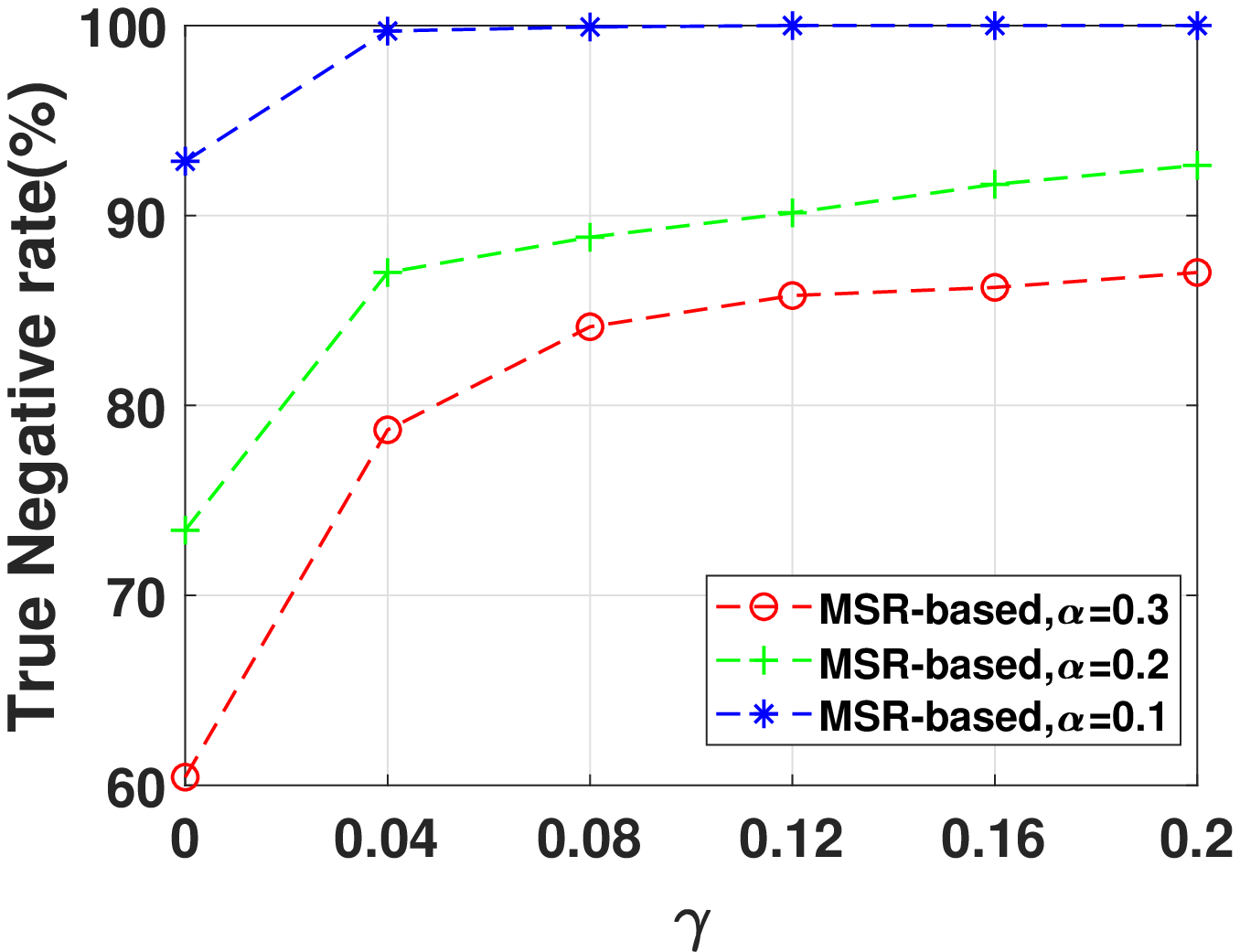}
 		\caption{}
 		\label{Gam_Efe_TNR}
 	\end{subfigure}
 	\begin{subfigure}[t]{0.3\linewidth}
 		\includegraphics[width=5cm,height=5cm]{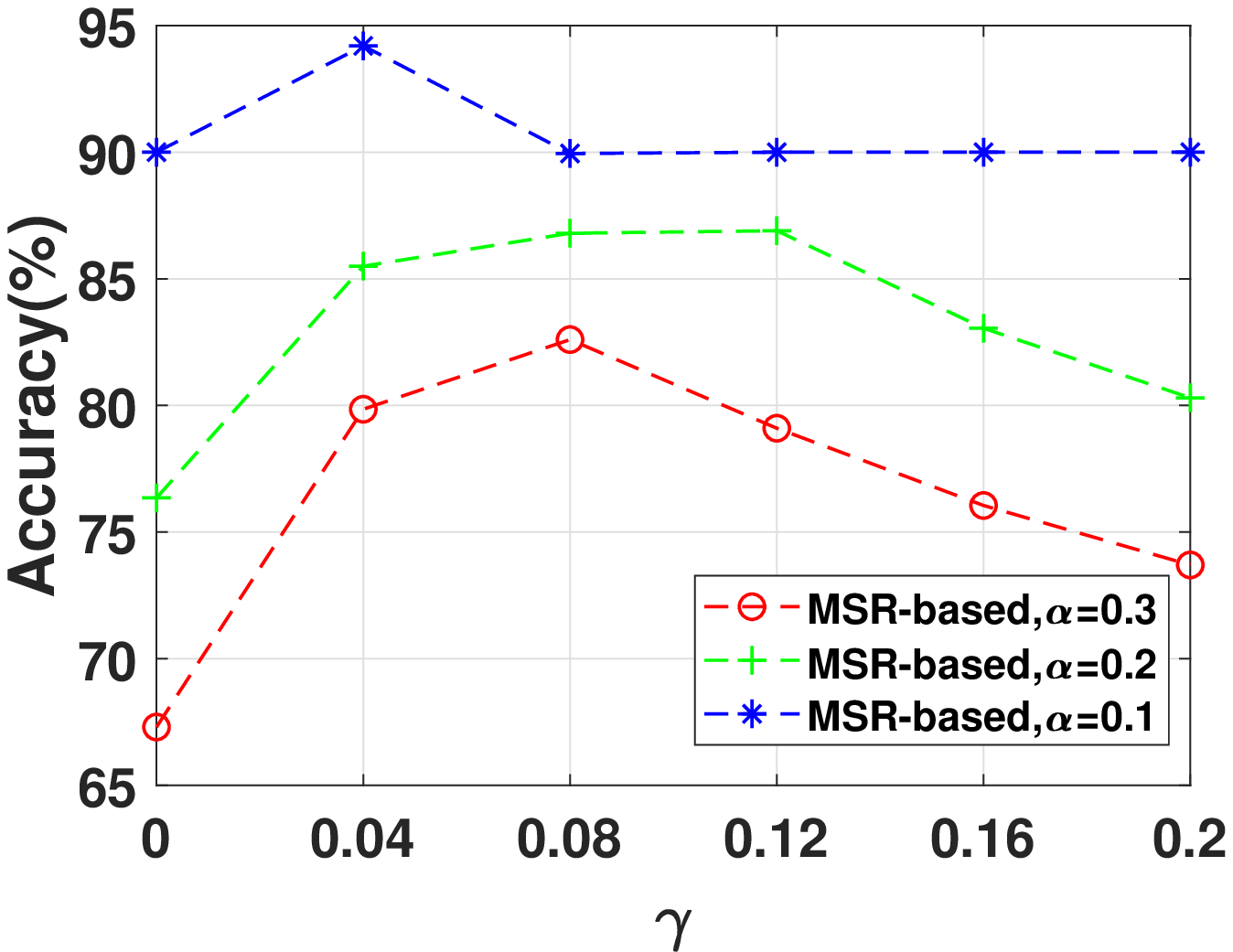}
 		\caption{}
 		\label{Gam_Efe_ACC}
 	\end{subfigure}
 	
 	\caption{(a) true positive rates, (b) true negative rates and (c) accuracies of the MSR-based algorithm with different $\gamma$ ranging from 0 to 0.2 in step of 0.04 when applied to the mixed AR data sequences at length 512 with different instantaneous coupling strength $\alpha=0.1, 0.2, 0.3$. The results are shown as an average over 100 realizations.}
 	\label{Gam_Efe}
 \end{figure*}

The non-linear AR data series were first generated for 100 realization at data lengths 512 using \eqref{AR} and then mixed using $\eqref{MIxed}$ in order to evaluate the effect of IC on the performance of the NUE algorithms. As it was mentioned in Sections 3, selecting a decent $\gamma$ can control the balance between true positives and true negatives which can be useful, for example, to increase the accuracy of our proposed MSR-based NUE algorithm when there is an unobserved confounder effect like IC effect. Therefore, the proposed algorithm was implemented using six $\gamma$s. We set a fixed $\lambda=0.5$ since in this section the goal is to investigate effect of $\gamma$ on the performance of the MSR-based algorithm. Figure \ref{Gam_Efe} demonstrates the TPRs, TNRs and accuracies of the MSR-based with six $\gamma$ when they are applied on the data with three instantaneous couplings, i.e $\alpha=0.1, 0.2, 0.3$. As we can see in Figure \ref{Gam_Efe}, the TNR of the MSR-based algorithm increases by increasing $\gamma$ while the TPR gradually decreases up to a certain $\gamma$ (e.g., $\gamma=0.04$ for $\alpha=0.1$) and then it significantly declines. Accordingly, the accuracy increases up to a certain $\gamma$ due to the increasing of the TNR compensating for the slight decrease of the TPR. Table \ref{Tab} illustrates accuracies of the existing NUE algorithms as well as the best accuracy of the MSR-based algorithm which is obtained by a reported $\gamma$ in the table. As Table \ref{Tab} demonstrates, accuracies of the NUE algorithms decrease by increasing instantaneous effect strength. Our proposed MSR-based NUE algorithm attains the greatest accuracy compared to the existing algorithms for all $\alpha$s.

 \begin{table}[H]
 	\caption{Accuracies of the bootstrap-based, LA-based and AIC-based algorithms as well as the proposed MSR-based algorithm. The $\gamma$ leads to the best accuracies of the MSR-based algorithm are also reported in parenthesis after the accuraies. The results are reported as an average over 100 realizations.}
 	\label{Tab}
 	\centering
 	\begin{tabular}{ccccc}
 		\toprule
 		
 		\textbf{NUE algorithm}	 &\textbf{Bootstrap-based} & \textbf{LA-based} & \textbf{AIC-based}&\textbf{MSR-based} (best $\gamma$)\\ 
 		\midrule
 		\textbf{$\alpha=0.1$}	&71.10&86.30&88.65&\textbf{94.20}($\gamma=0.04$)\\
 		\textbf{$\alpha=0.2$}	&71.80&77.75&73.20&\textbf{86.90}($\gamma=0.12$)\\
 		\textbf{$\alpha=0.3$}	&63.55&75.10&60.55&\textbf{82.60}($\gamma=0.08$)\\
 		\bottomrule
 	\end{tabular}
 \end{table}

 \subsubsection{Execution Time}
 In this section the AR model data at length 512 was generated and execution time of the NUE algorithms is reported as an average over 100 realizations. Similar to the results reported in Section 5.1.3, the MSR-based algorithm with $\lambda=1$ (at fixed $\gamma=0$) is the fastest algorithm and LA-based one is the slowest one. Although the total number of iterations of the AIC-based and MSR-based algorithms with $\lambda=0$ before termination are almost the same (around 10 iterations), the execution time of the AIC-based is slightly higher. It can be due to the fact that we did not have access to optimal code for calculating the KDE-based regression while for the NN-based prediction we have used a mex file for the neighbor search which is provided by the ITS toolbox \cite{faes2016information}. 
 \begin{table}[H]
 	\caption{The execution time of our proposed MSR-based with $\lambda=0,1$  (at fixed $\gamma=0$) as well as existing NUE algorithms for the AR data at length 512. The results are reported as an average over 100 realizations.}
 	\label{ET_HM}
 	\centering
 	\begin{tabular}{cccccc}
 		\toprule
 		
 		\textbf{NUE algorithm}  &\textbf{Bootstrap-based} & \textbf{LA-based} & \textbf{AIC-based}&\textbf{MSR-based, $\lambda=1$}&\textbf{MSR-based, $\lambda=0$}\\ 
 		\midrule
 		\textbf{Execution Time}	&28.96&38.02&5.09&1.62&3.64\\
 		\textbf{Total Number of Iterations}	&14.13&7.65&10.15&10.61&10.12\\
 		\bottomrule
 	\end{tabular}
 \end{table}

 \begin{figure*}[t]
 	\centering
 	\begin{subfigure}{0.3\linewidth}
 		\caption{Pre-Ictal (MSR-based)}
 		\label{MSR_Pre}
 		\vspace{0.1cm}
 		\includegraphics[width=5cm,height=5cm]{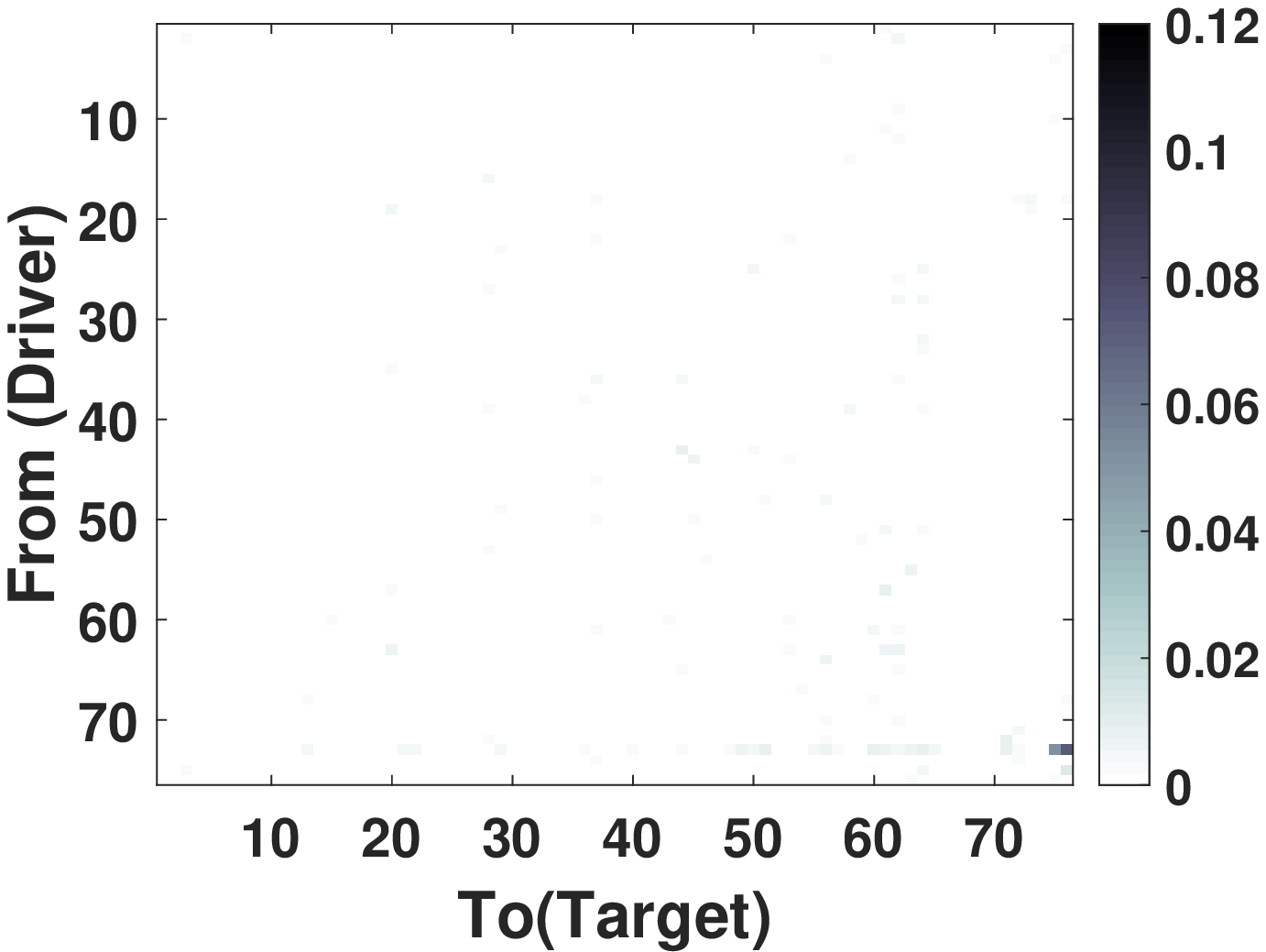}
 	\end{subfigure}
 	\begin{subfigure}{0.3\linewidth}
 		\caption{Pre-Ictal (Bootstrap-based)}
 		\label{Boot_Pre}
 		\vspace{0.1cm}
 		\includegraphics[width=5cm,height=5cm]{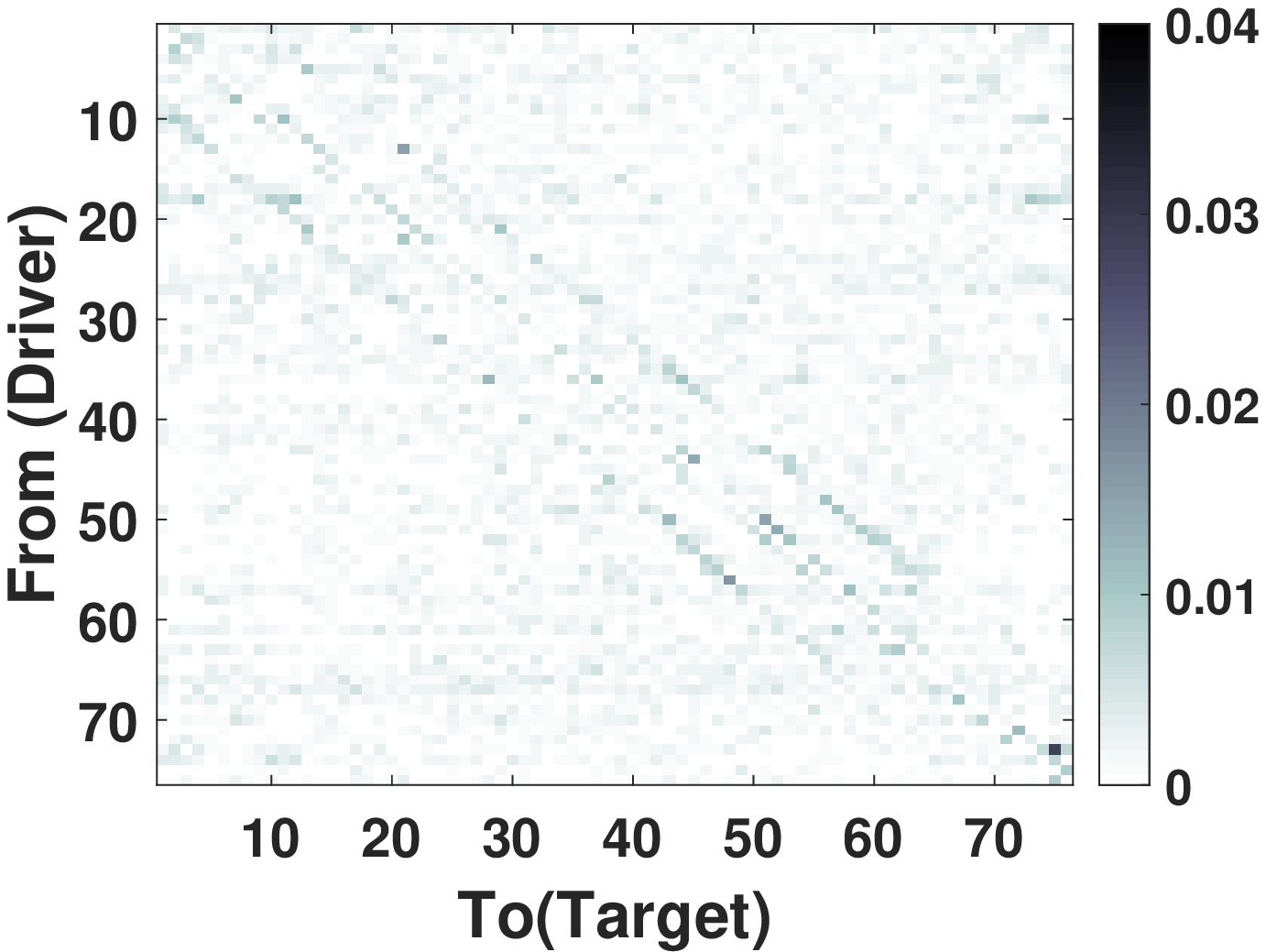}
 	\end{subfigure}
 	\begin{subfigure}{0.3\linewidth}
 		\caption{Pre-Ictal (LA-based)}
 		\label{LA_Pre}
 		\vspace{0.1cm}
 		\includegraphics[width=5cm,height=5cm]{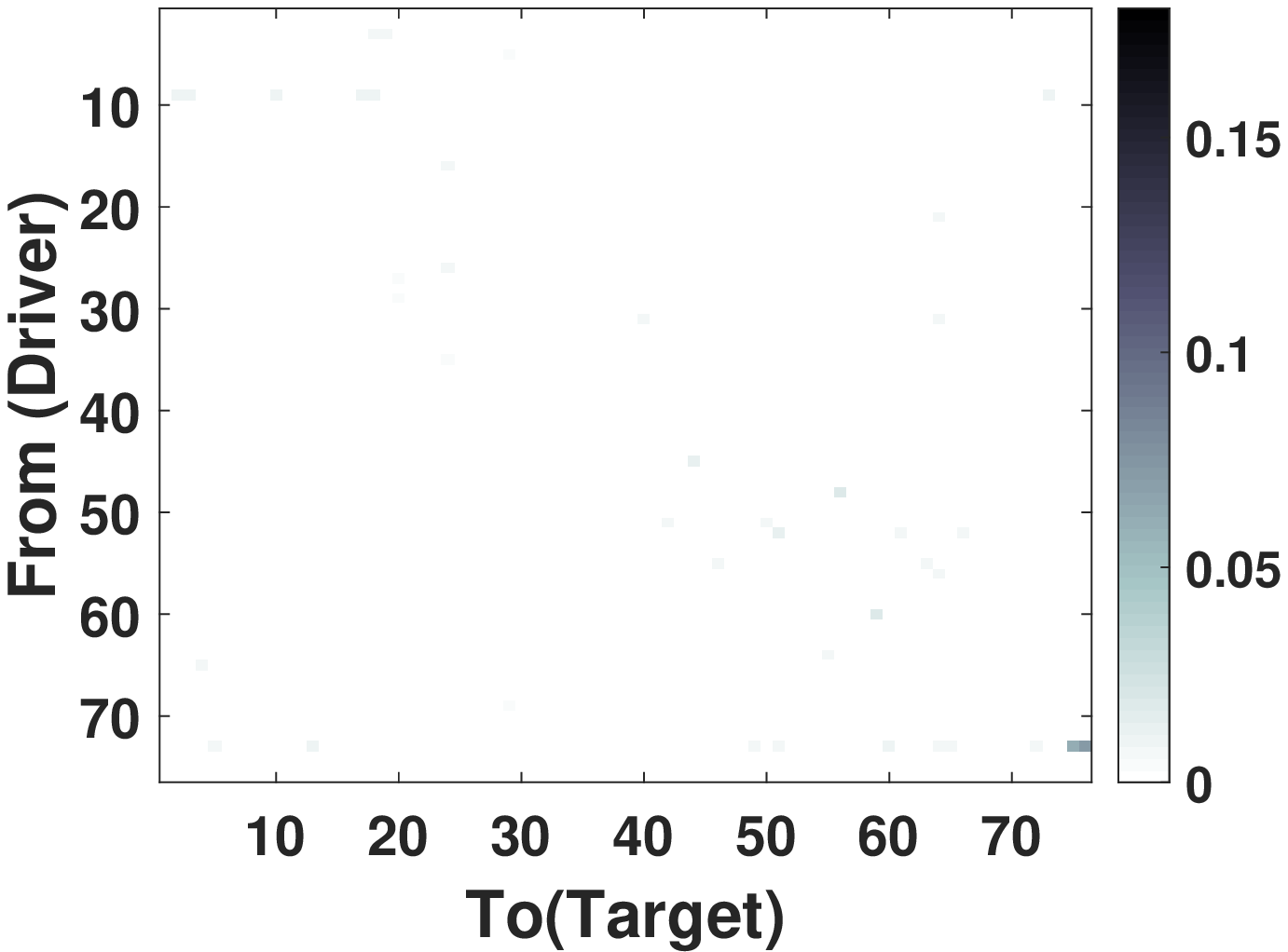}
 	\end{subfigure}
 	\begin{subfigure}{0.3\linewidth}
 		\caption{Ictal (MSR-based)}
 		\label{MSR_Ict}
 		\vspace{0.1cm}
 		\includegraphics[width=5cm,height=5cm]{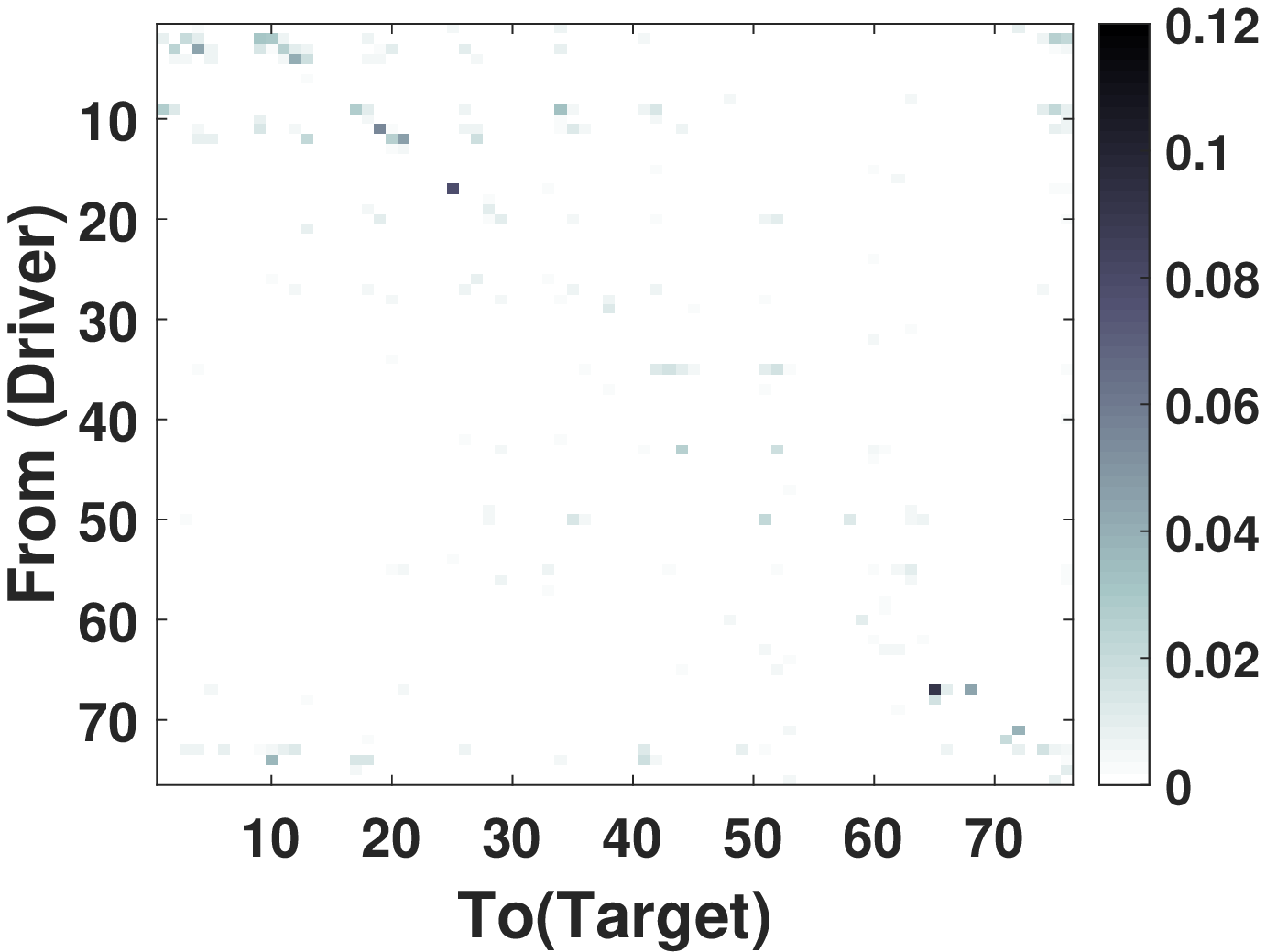}	
 	\end{subfigure}
 	\begin{subfigure}{0.3\linewidth}
 		\caption{Ictal (Bootstrap-based)}
 		\label{Boot_Ict}
 		\vspace{0.1cm}
 		\includegraphics[width=5cm,height=5cm]{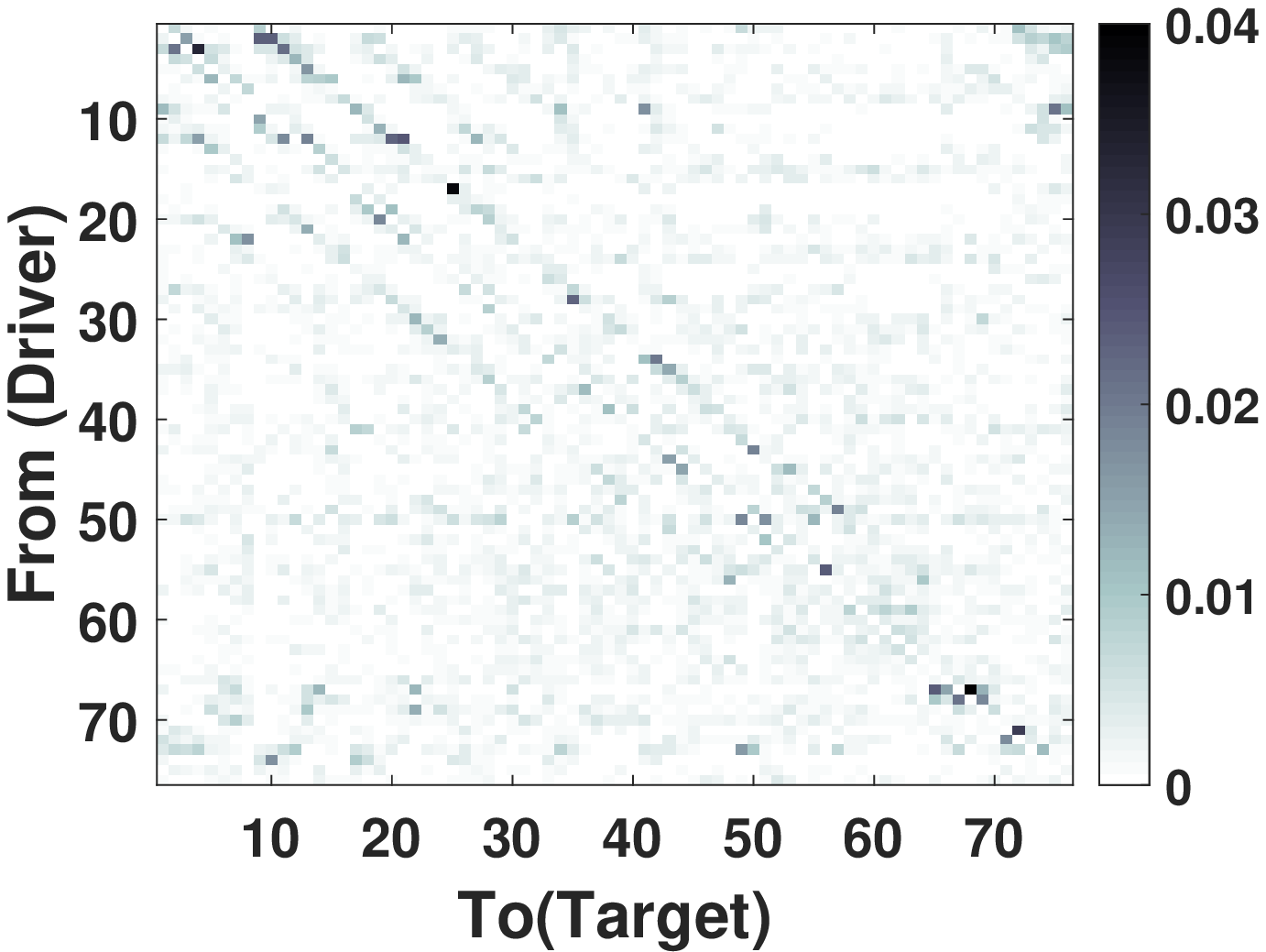}	
 	\end{subfigure}
 	\begin{subfigure}{0.3\linewidth}
 		\caption{Ictal (LA-based)}
 		\label{LA_Ict}
 		\vspace{0.1cm}
 		\includegraphics[width=5cm,height=5cm]{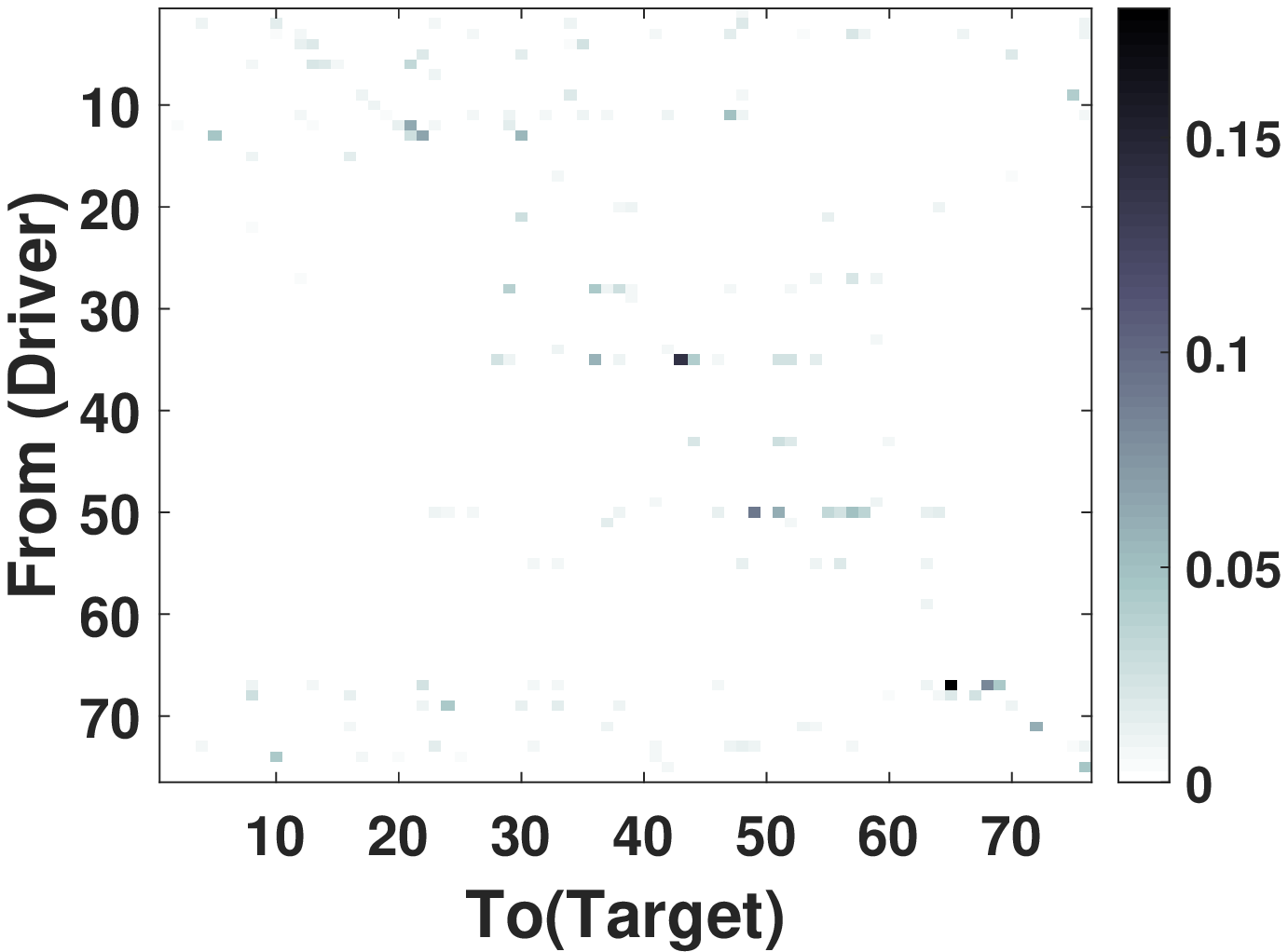}
 	\end{subfigure}
 	\caption{directed dependency Matrices obtained by applying NUE algorithms on intracranial EEG data at epileptic seizures (Ictal) and just before the seizure onset (Pre-ictal) conditions. The directed dependency is shown from rows (driver) to the colomns (Targets). The darker color of an element, the higher the directed dependency is. The results are shown as an average over 8 segments}
 	\label{CM EEG}
 \end{figure*}
\section{Application}
In this section, we demonstrate the applicability of our proposed MSR-based algorithm on a real-world data. We consider a publicly available high dimensional intracranial EEG data from an epileptic patient. While our proposed estimator is defined for stationary stochastic processes, at least for this particular case of real world EEG data, our estimator is also able to provide good results when applied on non-stationary signals. The overall goal here is to apply NUE algorithms to estimate CTE and find patterns related to the onset and spread of the seizure. A total of 76 implanted electrodes was recorded, resulting in 76 time series. Electrodes 1-64 are cortical electrode grid and electrodes 65-76 are in-depth electrodes (six electrodes on each side). The data comprises 8 epileptic seizures (Ictal) and 8 periods just before the seizure onset (Pre-ictal) segments. Each segment is 10 seconds intracranial EEG data recorded at 400 Hz sampling frequency (more details about this data can be found in \cite{kramer2008emergent}). In this work, an anti-aliasing low-pass filter with a cutoff frequency of 50 Hz was applied prior to downsampling the signals to 100 Hz \footnote{Slow temporal auto-correlation of signals can induce a bias in the estimated conditional TE , non-linear prediction and CMI in the NUE algorithms \cite{wibral2014directed}. An approach used to correct this bias is called Theiler correction based on which too close observations in time should be discarded from the NN searches included in the estimation of TE, CMI and MSR \cite{wibral2014directed}. In this paper we down-sample the EEG data to avoid slow auto-correlation bias. In other words, the Theiler window is 4 samples.}. The embedding delay and dimension were chosen as 1 and 8, respectively.  

Epileptologists recognized the regions corresponding to one of the depth strips (electrodes 70 to 76) and the lower left corner of the grid (electrodes 1–4, 9–11 and 17) were resected during anterior temporal lobectomy as the seizure onset zone, which means synchronous activity of neurons in the specific regions of the brain becomes so strong, so that it can propagate its own activity to other distant regions \cite{jia2019detecting,montalto2014mute,faes2017multiscale,zhang2018low}. From an information theory point of view, these nodes send information to other nodes, resulting in seizure onset. The amount of information each node sends to other nodes can be computed by the summation over each row of the directed dependencies matrix. 

We applied our proposed in addition to bootstrap-based and LA-based NUE algorithms to estimate CTE in real high dimensional and redundant intracranial EEG data. The overall goal here is to compare advantages of our proposed NUE algorithms over the other algorithms reported in the literature. The MSR-based NUE algorithm was implemented with $\lambda=1$ and $\gamma=0.005$. The directed dependencies matrices obtained by our proposed algorithm as well as the existing algorithms are shown in Figure \ref{CM EEG}. The directed dependencies matrices obtained by the bootstrap-based NUE algorithm (Figures \ref{Boot_Ict} and \ref{Boot_Pre}) contain many connections in both pre-ictal and ictal conditions. Specifically, the diagonal pattern observed in the matrices obtained by the bootstrap-based NUE algorithm can be due to the volume conduction and conduction effect of the grid. On the other hand, our proposed (Figures \ref{MSR_Ict} and \ref{MSR_Pre}) and LA-based NUE algorithms (Figure \ref{LA_Ict} and \ref{LA_Pre}) are less sensitive to the volume conduction effect in comparison to that of the bootstrap-based algorithm. 
        
\begin{figure*}[t]
	\centering
	\begin{subfigure}{0.3\linewidth}
		\caption{Pre-Ictal (MSR-based)}
		\label{Sending_Pre_MSR}
		\vspace{0.1cm}
		\includegraphics[width=5cm,height=5cm]{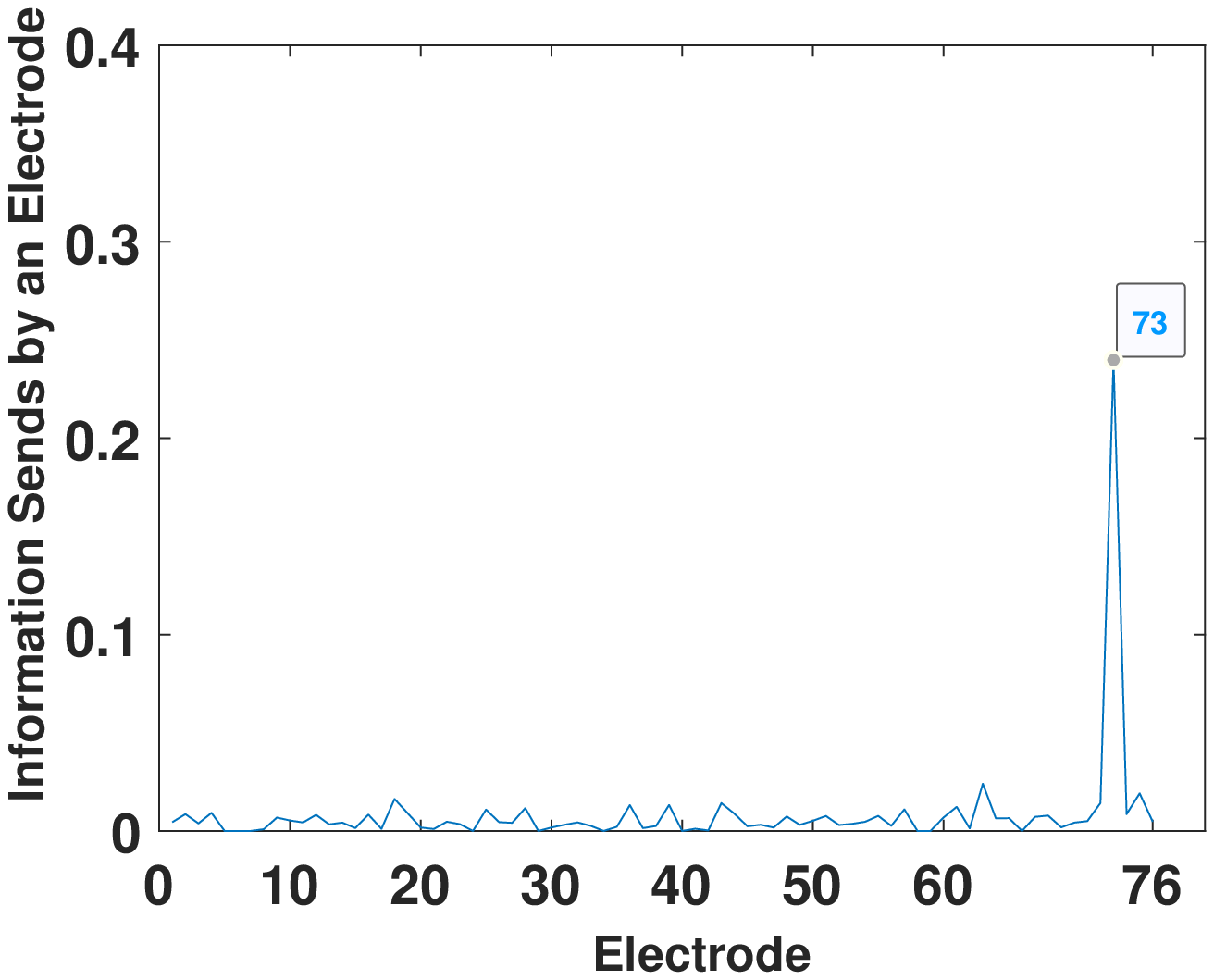}
	\end{subfigure}
	\begin{subfigure}{0.3\linewidth}
		\caption{Pre-Ictal (Bootstrap-based)}
		\label{Sending_Pre_Boot}
		\vspace{0.1cm}
		\includegraphics[width=5cm,height=5cm]{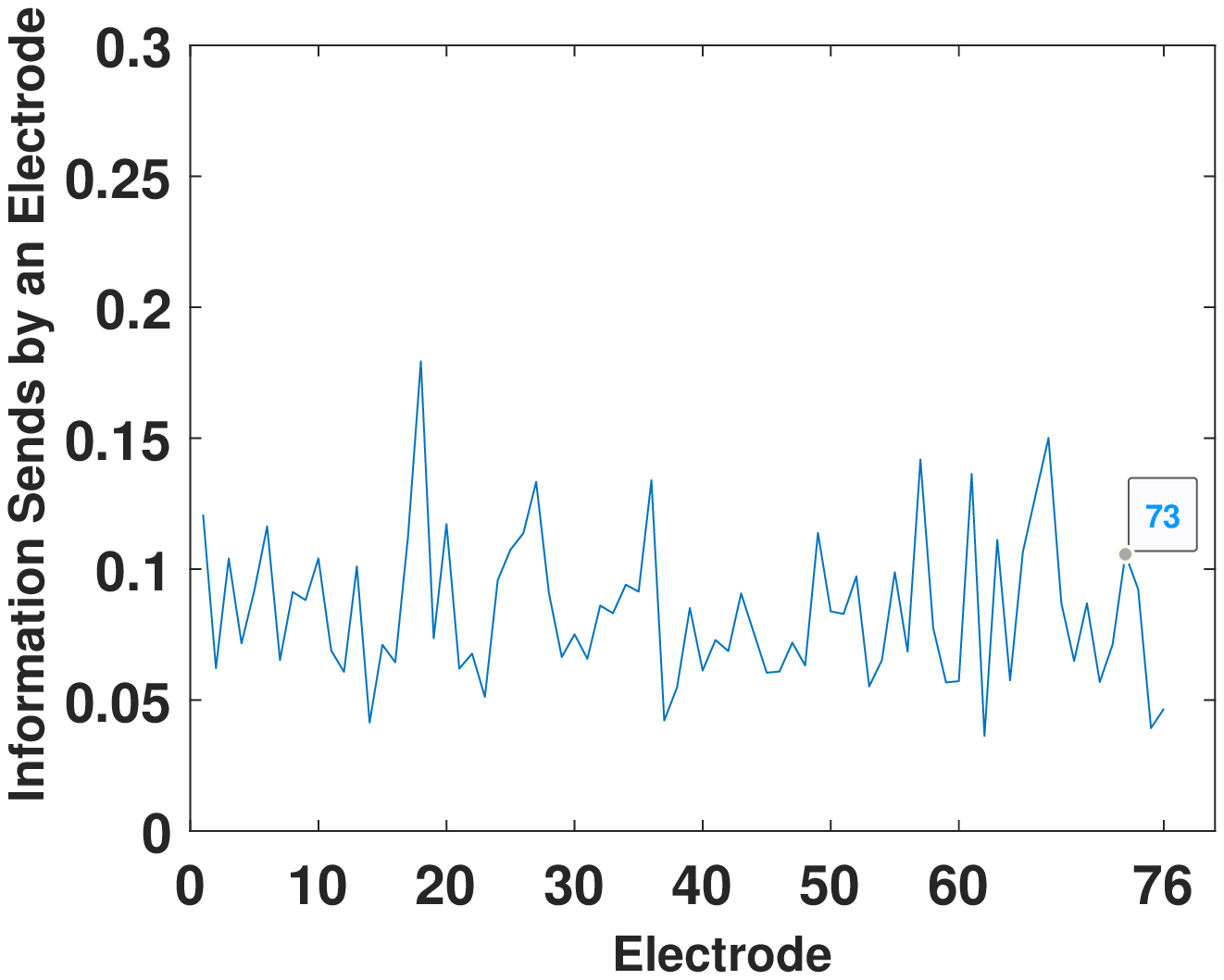}
	\end{subfigure}
	\begin{subfigure}{0.3\linewidth}
		\caption{Pre-Ictal (LA-based)}
		\label{Sending_Pre_LA}
		\vspace{0.1cm}
		\includegraphics[width=5cm,height=5cm]{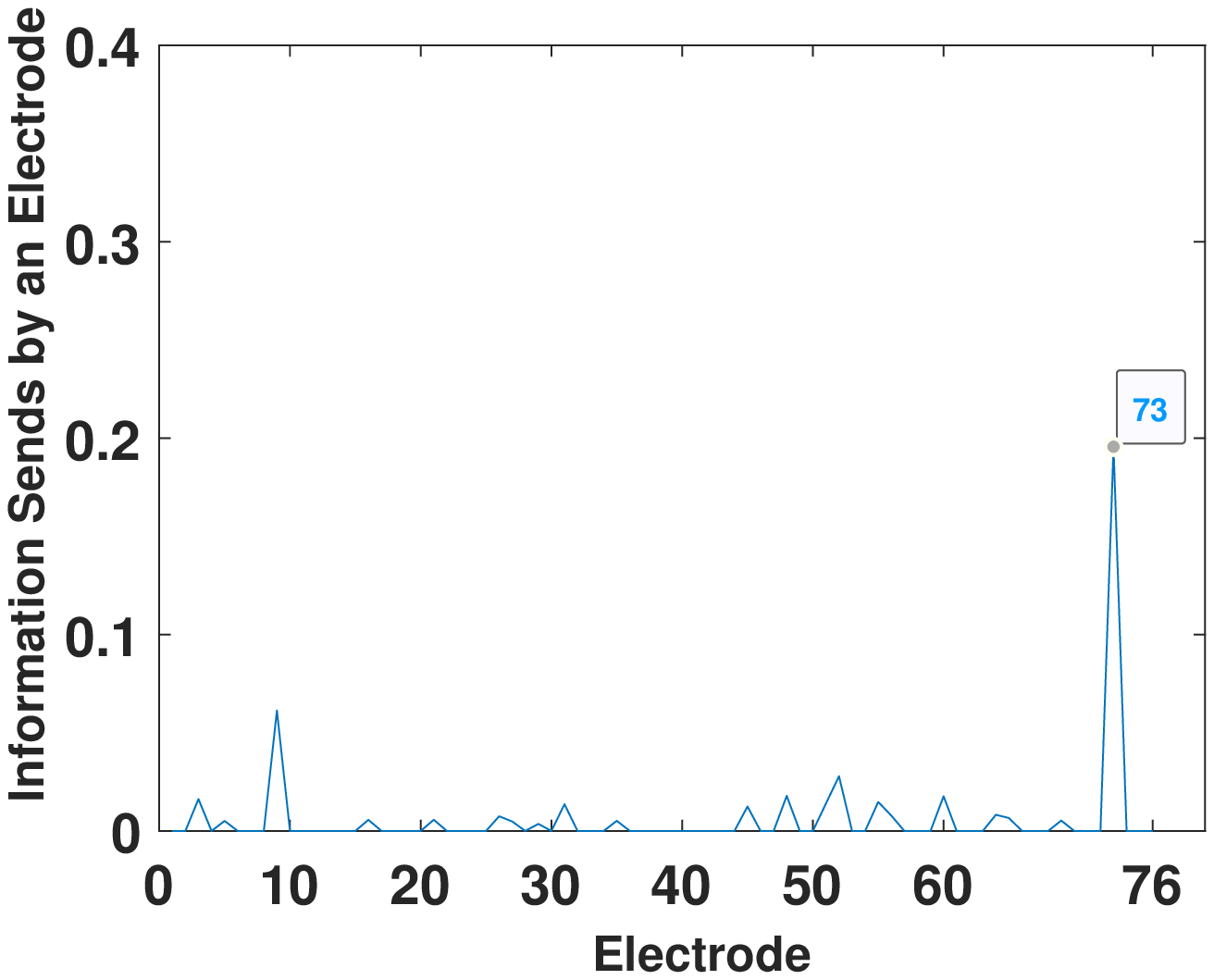}
	\end{subfigure}
	\begin{subfigure}{0.3\linewidth}
		\caption{Ictal (MSR-based)}
		\label{Sending_Ict_MSR}
		\vspace{0.1cm}
		\includegraphics[width=5cm,height=5cm]{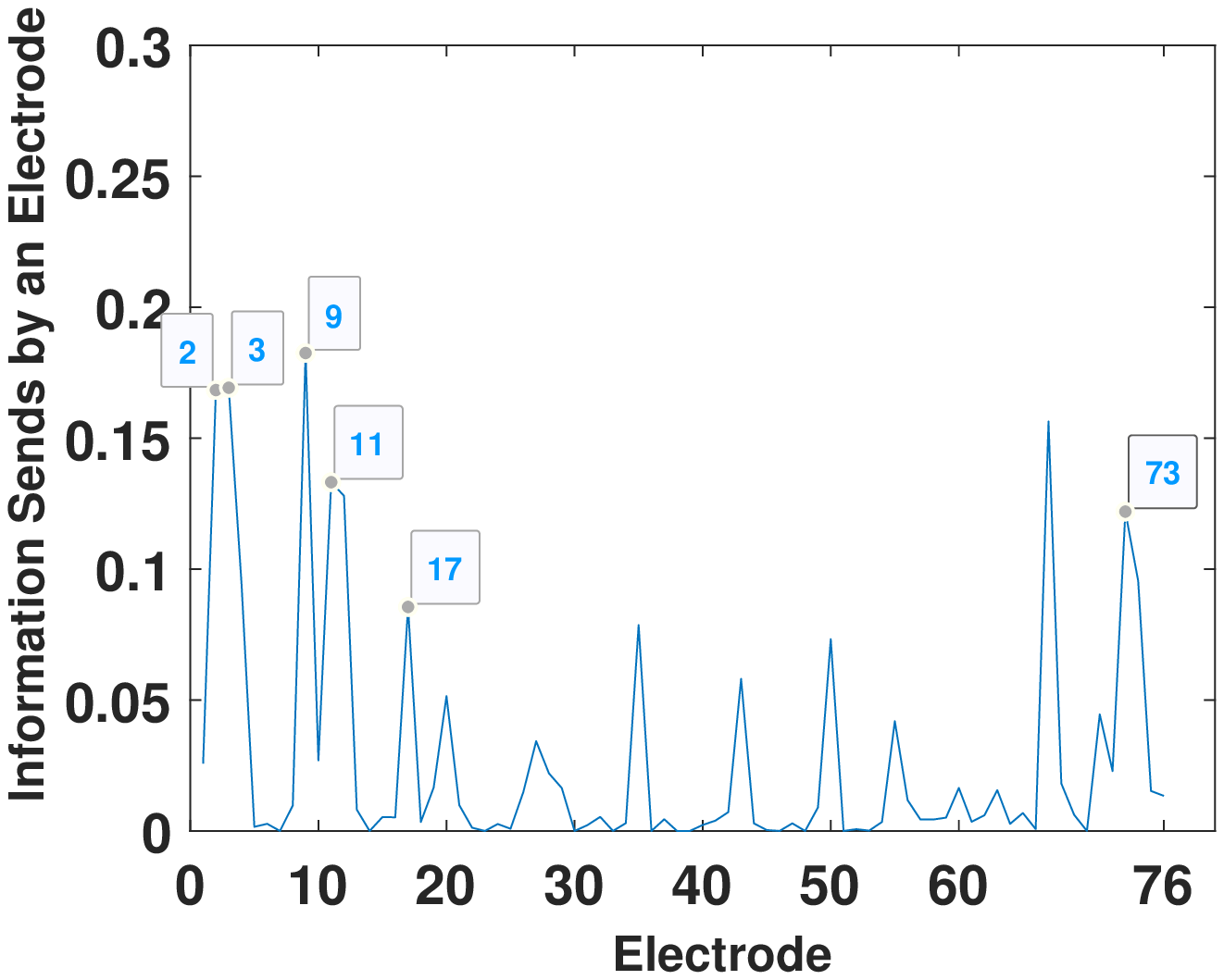}	
	\end{subfigure}
	\begin{subfigure}{0.3\linewidth}
		\caption{Ictal (Bootstrap-based)}
		\label{Sending_Ict_Boot}
		\vspace{0.1cm}
		\includegraphics[width=5cm,height=5cm]{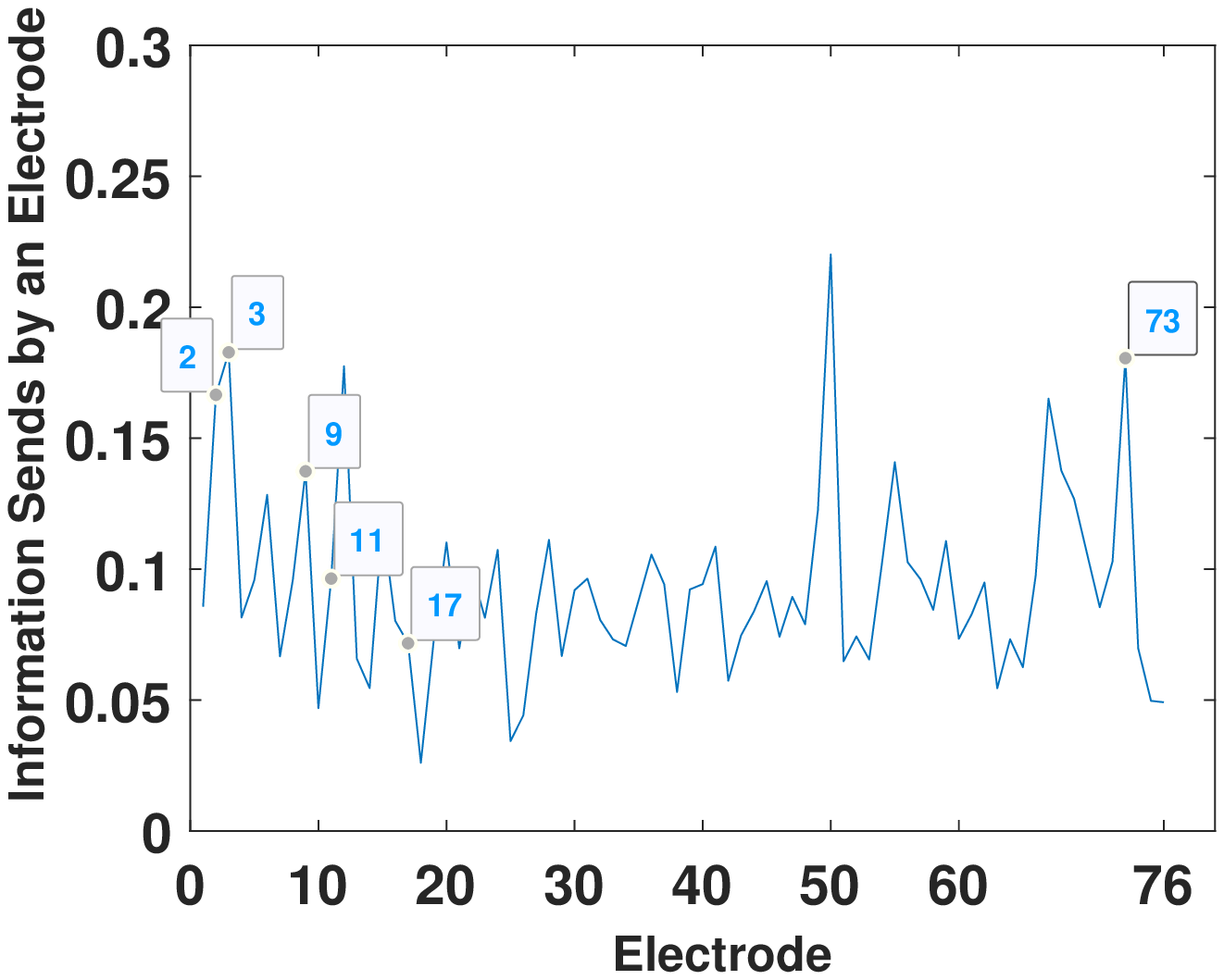}	
	\end{subfigure}
	\begin{subfigure}{0.3\linewidth}
		\caption{Ictal (LA-based)}
		\label{Sending_Ict_LA}
		\vspace{0.1cm}
		\includegraphics[width=5cm,height=5cm]{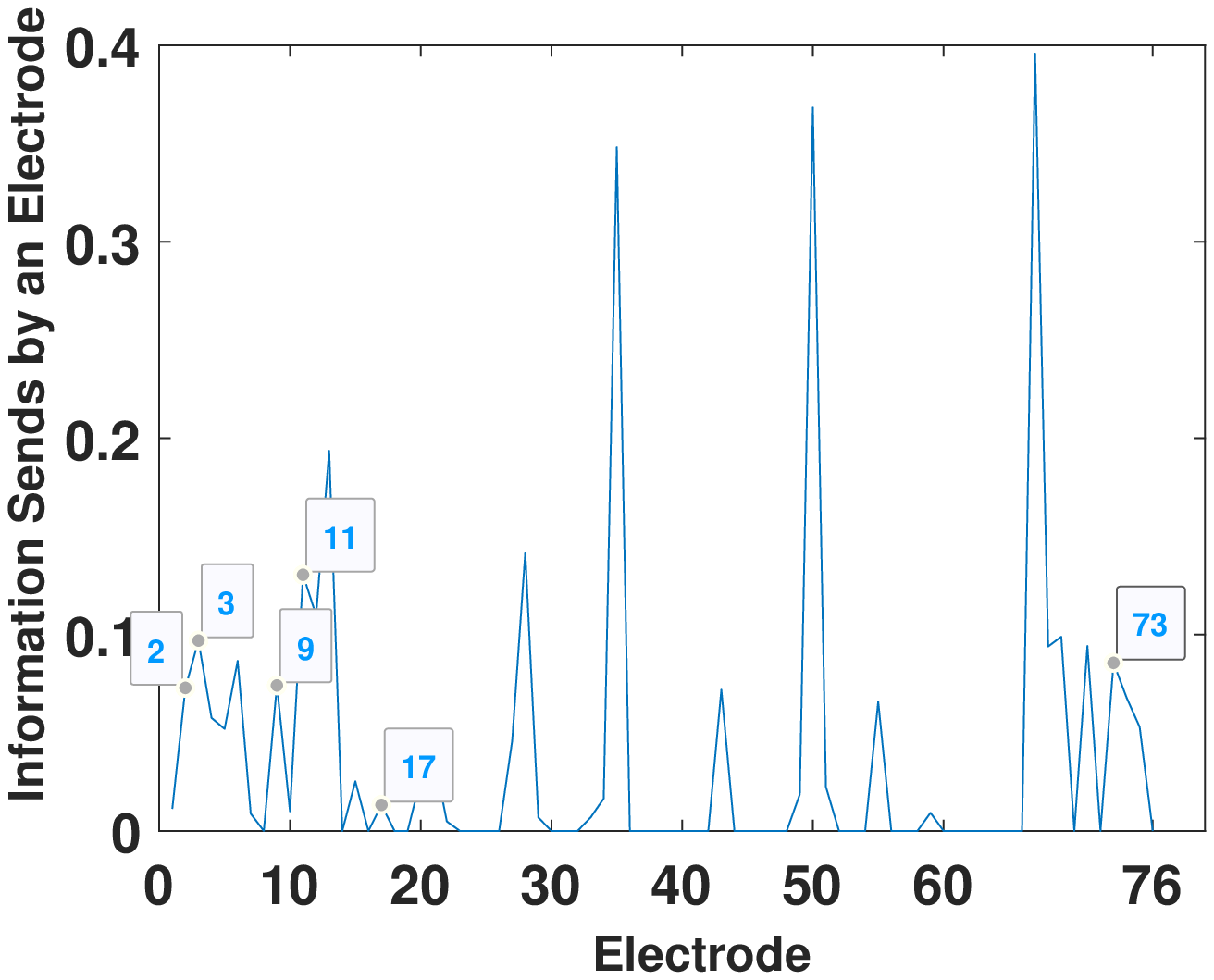}
	\end{subfigure}
\caption{Total Information each electrode sends to other contacts at ictal and pre-ictal conditions}
\label{Sending}
\end{figure*}

Figure \ref{Sending} represents the total amount of information each electrode sends to other electrodes. As Figure \ref{Sending_Pre_Boot} demonstrates, due to the volume conduction effect there are some peaks even in the pre-ictal condition. On the other hand, the amount of information each electrode sends in the pre-ictal condition obtained by the MSR-based (Figure \ref{Sending_Pre_MSR}) and LA-based (Figure \ref{Sending_Pre_LA}) NUE algorithms is approximately zero except for electrode 73. This electrode can be associated to the seizure onset although it is not yet clinically observable.

As mentioned earlier, electrodes 2-4, 9-11 and 17 are the seizure onset zones. Figure \ref{Sending_Ict_LA} and \ref{Sending_Ict_MSR} show that the magnitude of the peaks at electrodes 2-4 and 9-11 for the MSR-based algorithm is higher than the one of the LA-based procedure. It is also important to mention that the existing LA-based and bootstrap-based NUE algorithms are not able to detect the peak at electrode 17 as opposed to that of our proposed MSR-based NUE algorithm.   





\section{Discussion and Conclusion}
Reliable estimation of the directed dependencies in conditional high dimensional data is limited by the so-called ´´curse of dimensionality'' problem. A greedy approach called non-uniform embedding (NUE) algorithm was proposed in \cite{montalto2014mute} to select the most relevant variables and reduce the dimension of the reconstructed state-space of the data. Then, the model-free directed dependencies measure, conditional transfer entropy (CTE) is estimated using the reconstructed state-space. The NUE strategy based on sequentially selecting the best candidates in a greedy way will generally not lead to the same performance as would be obtained by using a brute-force combinatorial approach, where the performance is maximized over all possible sets of candidates. It has, however, been shown that NUE approaches often lead to an improved accuracy of the CTE compared to that of uniform embedding approaches \cite{faes2016information,montalto2014mute}. The NUE algorithm has been widely utilized to estimate the directed dependencies in neuro-physiological \cite{lindner2011trentool,wibral2013measuring} and economical \cite{bossomaier2016introduction} applications. It still has some obstacles like using a bootstrap-based termination criterion which highly depends on the bootstrap size \cite{may2008non}. It has been shown in \cite{massey1990causality,li2015improved} that using an alternative to the bootstrap statistical test can be more accurate and computationally efficient. 

In this paper, we proposed a new modification for the NUE algorithm which uses a weighted sum of conditional mutual information (CMI) and nearest neighbor(NN)-based prediction for ranking the candidates and the algorithm is terminated if the highest ranked candidate is not relevant enough to significantly improve the accuracy of the prediction of the target variable. It should be noted that while our simulations on synthetic and real world data indicate that using prediction accuracy can lead to better assessment of directed dependency, we have not been able to prove this from an estimation theoretic point of view. It should also be noted that for the linear Gaussian processes, accuracy of the prediction of the target variable given selected candidates $\operatorname{MSR}(Y_n|\mathcal{U}_n)$, is monotonically equivalent to the conditional entropy $H(Y_n|\mathcal{U}_n)$ \cite{barnett2009granger}.

The proposed NUE procedure was compared with the original bootstrap-based NUE algorithm in \cite{montalto2014mute}, low-dimensional approximation(LA)-based \cite{zhang2018low} and Akaike information criterion (AIC)-based \cite{may2008non}. Performance analysis using simulation data generated by Henon map and autoregressive (AR) models at different lengths and coupling strengths revealed that the proposed mean of the squared (MSR)-based NUE algorithm tends to outperform the existing ones for detecting the directed dependencies. Specifically, the higher true negative rate (TNR) of the proposed MSR-based NUE compared to that of the existing ones may represent better ability of the proposed algorithm to terminate at the correct iteration and as a result better functionality of the proposed termination criterion. The poor selectivity (or TNR) of the bootstrap-based is in line with the results observed in \cite{zhang2018low}, where they also found higher false positive for the bootstrap-based procedure compared to that of the LA-based one. The proposed algorithm also attains less false positive in comparison to that of the LA-based approach. The greater true positive rate (TPR) of the MSR-based algorithm with higher $\lambda$ for small simulated data length and low coupling strength can justify using the weighted sum for ranking candidates. However, the limitation of the proposed NUE algorithm is that for very low coupling strength, the accuracy of the proposed estimator was not as good as for the the bootstrap-based one.

The applicability of the NUE algorithms in real-word data can be affected by unobserved confunder effects like instantaneous information sharing which can be falsely detected as directed dependencies \cite{faes2013compensated}. The data sequences generated by AR model were instantly mixed at different mixing strength in order to simulate instantaneous coupling (IC) effect. The results showed that by choosing a proper parameter $\gamma$, the proposed MSR-based measure attains significantly better performance than the existing ones. The simulated data results were consistent with the real-data used in this paper where the best results also were obtained for positive $\gamma$. The better performance can be of particular importance for such real-world application like electroencephalography (EEG) and magnetoencephalography in which the volume conduction effect can cause IC \cite{ruiz2019computational}. There are also other frameworks like compensated transfer entropy  \cite{faes2013compensated} which tries to improve the estimation of the TE in presence of IC. This measure modified the definition of the transfer entropy to compensate the effect of IC. The NUE algorithms are defined to find the embedding vector for estimating transfer entropy. Therefore, comparison or even modification of the proposed NUE algorithm for restructuring the state-space to estimate compensated transfer entropy deserves an independent and comprehensive study and will be considered in future works.     

The proposed MSR-based algorithm with known parameter $\gamma$ achieved a significant improvement in the computational efficiency. This can be due to the elimination of the computation effort of the bootstrap test which is not included in the proposed MSR-based algorithm. If we consider that the estimation of the CMI dominates the computation of the NUE algorithms (except for the MSR-based with $\lambda=1$) then the overall computational requirement of the NUE algorithms which uses bootstrap-based test in the worst case will be $k|\mathcal{C}|+100k$, where $k$ is the number iterations reported in Tables 1 and 2 and $|\mathcal{C}|$ is the cardinality of $\mathcal{C}$. On the other hand, the computational requirement of the proposed NUE algorithm with known parameter $\gamma$ can be expressed as $k|\mathcal{C}|$. The computational effort of the MSR-based NUE algorithm with $\lambda=1$ can be considered to be dominated by the estimation of MSR. It is computationally less complex than that of the CMI since it only includes a neighbor search while CMI estimation contains a neighbor search and range searches. Therefore, in very high-dimensional data (like the intracranial EEG data used in the application part where $|\mathcal{C}|=608$ ) where execution of the NUE algorithm can be very time-consuming, it is suggested to use $\lambda=1$ since it will be significantly faster. The proposed NUE algorithm with $\lambda=1$ also achieved better execution time than that of with $\lambda=0$ in the simulation data used this paper. As already mentioned in \cite{jia2019detecting,zhang2018low}, the LA-based approximation of the CMI used in the LA-based NUE algorithm is computationally more expensive and this is consistent with the execution time reported in this paper where LA-based procedure attains the worst execution time. Better execution time can be especially important for such applications like scalp EEG-based brain-computer interface where faster time series analyses methods are required. We also consider to test the performance our proposed estimator on high dimensional scalp EEG data in future works.

Another parameter of our proposed NUE algorithm over which one needs to scan is the positive parameter $\gamma$. The parameter $\gamma$ and $\operatorname{MSR(Y_n|\mathcal{U_n})}$ have the same units and it defines the required amount of improvement in the accuracy of prediction prior to selecting a variable. Intuitively, the prediction accuracy $MSR(Y_n|\mathcal{U}_n^k)$ can vary between 0 and $\operatorname{var}(y_n)$ where 0 shows that one can perfectly predict $Y_n$ by incorporating $\mathcal{U}_n^k$. The intuition of the worst case of the accuracy of the prediction $MSR(Y_n|\mathcal{U}_n^k)$ can be the case that incorporating $\mathcal{U}_n^k$ does not help the prediction at all and the indices specified by neighbor search in $\mathcal{u}_n$ will be uniformly distributed. The obtained $\widehat{y_n}(i|\mathcal{U})$ will be an approximation of mean of $y_n$ and as a result $\operatorname{MSR}$ will be approximately $\operatorname{var}(y_n)$. In this paper, we normalize time series related to the realizations of the target processes to have zero mean and unit variance. We therefore scan the parameter $\gamma$ in the interval between 0 and 1 to tune the algorithm. Therefore, another limitation of our proposed NUE algorithm is that it needs to be tuned by scanning over the parameter $\gamma$. The optimal choice of $\gamma$ will be data dependent. The more accurate investigation of the criterion with which the parameter $\gamma$ can be selected will be considered in the future works. Moreover, scanning over $\gamma$ can increase execution time of our proposed algorithm. We suggest tuning the algorithm by using small subset of segments and use the tuned algorithm for the rest of segments. The reason is that the parameter $\gamma$ can take care of confounder effects found in the data and will not vary during the segments such as volume conduction effect in neuro-physiological time series \cite{ruiz2019computational}.         
  
In this paper, TE has been used to assess the directed dependencies. Estimated TE in networks consisting of more than two nodes can be affected by other nodes through, for example, indirect path or common shared information. One possible approach to reduce such effects is to condition out information coming from other nodes. However, this approach can present bias in the estimated directed dependencies in data in which there is the collider condition \cite{cole2010illustrating}. There are other approaches to assess directed dependencies in the network like decomposing TE into unique, synergistic, and redundant information \cite{williams2011generalized}. However, comparison of the estimated conditional TE and decomposing TE deserves an independent and comprehensive study and it is out of scopes of this paper.

\vspace{6pt} 






\conflictsofinterest{``The authors declare no conflict of interest.'' .} 

%

\appendixtitles{no} 
\appendix
\section{}
\unskip
\subsection{}
\label{NN}
The Kraskov-Grassberger-St\"{o}gbauer approach \cite{kraskov2004estimating} is an NN-based estimator which was originally developed in order to estimate mutual information. It was adapted to estimate CMI in the NUE algorithm in \cite{montalto2014mute,faes2015estimating,faes2016information}. The CMI in \eqref{Select} can be rewritten as the sum/difference of four joint entropies \cite{faes2015estimating,montalto2014mute}
\begin{equation}
\operatorname{I}\left(Y_n;W_n\mid{\mathcal{S}_n^{k-1}}\right)=h(Y_n,\mathcal{S}_n^{k-1})-h(\mathcal{S}_n^{k-1})
-h(Y_n,W_n,\mathcal{S}_n^{k-1})+h(W_n,\mathcal{S}_n^{k-1}).
\label{Entropies}
\end{equation}   

Then, the CMI  is estimated by using a NN approach in which the entropy of the higher dimension $h(Y_n,W_n,\mathcal{S}_n^{k-1})$ is estimated through a neighbor search as \cite{faes2015estimating,montalto2014mute} 
\begin{equation}
h(Y_n,W_n,\mathcal{S}_n^{k-1})\approx -\psi (T)+\psi (N)+(d+1)\langle ln(\epsilon_n(i)\rangle),
\label{NS}
\end{equation}
where $\psi$ is the digamma function and $N$ is total the number of observations of the vector variable $[Y_n,W_n,\mathcal{S}_n^{k-1}]$. Twice distance (maximum norm) of $i^{th}$ observation of $[Y_n,W_n,\mathcal{S}_n^{k-1}]$ from its $T^{th}$ neighbor is denoted by $\epsilon_n(i)$ and $\langle . \rangle$ is the average over all observations. The rest of entropies in \eqref{Entropies} are estimated by using a range search as 
\begin{equation}
\begin{aligned}
h(W_n,\mathcal{S}_n^{k-1})\approx -\psi (T)+d\left\langle \psi \left(N_{[W_n,\mathcal{S}_n^{k-1}]}+1\right)\right\rangle\\
h(Y_n,\mathcal{S}_n^{k-1})\approx -\psi (T)+d\left\langle \psi \left(N_{[Y_n,\mathcal{S}_n^{k-1}]}+1\right)\right\rangle\\
h(\mathcal{S}_n^{k-1})\approx -\psi (T)+(d-1)\left\langle \psi \left(N_{\mathcal{S}_n^{k-1}}+1\right)\right\rangle.
\end{aligned}
\label{rS}
\end{equation}

The number of realizations of $[W_n,\mathcal{S}_n^{k-1}]$ whose maximum norm from the $i^{th}$ realization of $[W_n,\mathcal{S}_n^{k-1}]$ is strictly less than $\epsilon_n/2$, is denoted by $N_{[W_n,\mathcal{S}_n^{k-1}]}$. A similar notation applies to $N_{[Y_n,\mathcal{S}_n^{k-1}]}$ and $N_{\mathcal{S}_n^{k-1}}$. The CMI is finally estimated by replacing \eqref{NS} and \eqref{rS} in \eqref{Entropies}
\begin{equation}
I\left(Y_n;W_n\mid{\mathcal{S}_n^{k-1}}\right)=\psi(T)+\left\langle\psi\left(N_{\mathcal{S}_n^{k-1}}+1\right)
-\psi\left(N_{[W_n,\mathcal{S}_n^{k-1}]}+1\right)-\psi\left(N_{[Y_n,\mathcal{S}_n^{k-1}]}+1\right)\right\rangle.
\end{equation} 

\subsection{}
\label{NNPTE}
After selecting the most informative candidates and forming the embedding vector $\mathcal{S}_n^k$ using the NUE algorithms, the CTE in \eqref{PTE} can be estimated using the same approach explained in \ref{NN}. The CTE can also be expressed as the sum of four joint entropies as
\begin{equation}
\operatorname{CTE}(\mathcal{X}\rightarrow \mathcal{Y}|\mathbfcal{Z})= h(Y_n,Y_n^-,\textbf{Z}_n^-)-h(Y_n^-,\textbf{Z}_n^-)-h(Y_n,Y_n^-,X_n^-,\textbf{Z}_n^-)+h(Y_n^-,X_n^-,\textbf{Z}_n^-),
\end{equation}   
where $[X_n^-,Y_n^-,\textbf{Z}_n^-]$ is replaced by $\mathcal{S}_n^{k-1}$ and $[Y_n^-,\textbf{Z}_n^-]$ is substituted by $\mathcal{S}_n^{k-1}$ without any past variables of $X_n$. Then by using range search in the higher dimension $[Y_n,Y_n^-,\textbf{Z}_n^-]$ and range search in the rest of dimensions, the CTE can be estimated as
\begin{equation}
\operatorname{CTE}(\mathcal{X}\rightarrow \mathcal{Y}|\mathbfcal{Z})=\psi(T)+\left\langle\psi\left(N_{[Y_n^-,\textbf{Z}_n^-]}+1\right)
-\psi\left(N_{[Y_n,Y_n^-,\textbf{Z}_n^-]}+1\right)-\psi\left(N_{[X_n^-,Y_n^-,\textbf{Z}_n^-]}+1\right)\right\rangle,
\end{equation}
where $N_{[Y_n^-,\textbf{Z}_n^-]}$ denotes the number of realizations of whose maximum norm from its $i^{th}$ realization of is strictly less than $\epsilon_n/2$. The same notation applies to $N_{[Y_n,Y_n^-,\textbf{Z}_n^-]}$ and $N_{[X_n^-,Y_n^-,\textbf{Z}_n^-]}$.

\section{}
\label{KDE}
In the Akaike information criterion-based termination criterion which is adapted in this paper to stop the NUE algorithm one needs to predict the target variable $Y_n$ given $\mathcal{U}_n^k=[W_n^k,\mathcal{S}_n^{k-1}]$ by using the kernel density estimation (KDE) approach. The KDE-based prediction is performed as:
\begin{equation}
\widehat{y}_n(i|\mathcal{U}_n^k)=\displaystyle\sum_{i=1}^N\frac{ \hspace{0.1cm}y_n(i)K_h(\mathcal{u}_n^k,\mathcal{u}_n^k(i))}{\sum_{i=1}^MK_h(\mathcal{u}_n^k,\mathcal{u}_n^k(i))},
\label{PMI}
\end{equation}
where $\widehat{y}_n(i|\mathcal{U}_n^k)$ denotes for estimated $i^{th}$ observation of $Y_n$ and $K_h$ is the Gaussian kernel with Mahalonobis distance (Eq. (\ref{Mahal}))  \cite{may2008non}:
\begin{equation}
K_h(\mathcal{u}_n^k,\mathcal{u}_n^k(i))=\frac{1}{(\sqrt{2\pi}h)^d}\exp\left(\frac{-\|\ \mathcal{u}_n^k-\mathcal{u}_n^k(i)\|}{2h^2}\right),
\label{Ker}
\end{equation}
\begin{equation}
\|\mathcal{u}_n^k-\mathcal{u}_n^k(i)\|=(\mathcal{u}_n^k-\mathcal{u}_n^k(i))^T\Sigma^{-1}(\mathcal{u}_n^k-\mathcal{u}_n^k(i)),
\label{Mahal}
\end{equation}
where $d$ and $\Sigma$ are dimension (number of columns) and covariance of $\mathcal{u}_n^k$, respectively. The bandwidth of the kernel function $h$ is chosen for unit variance data as \cite{may2008non,li2015improved}:
\begin{equation}
h=1.5\left(\frac{1}{d+2}\right)^{1/(d+4)}N^{-1/(d+4)}.
\end{equation}

\reftitle{References}




\bibliography{MyBib}

\begin{thebibliography}{-------}
\providecommand{\natexlab}[1]{#1}

\bibitem[Baboukani \em{et~al.}(2021)Baboukani, Graversen, and
  {\O}stergaard]{My}
Baboukani, P.S.; Graversen, C.; {\O}stergaard, J.
\newblock Estimation of Directed Dependencies in Time Series Using Conditional
  Mutual Information and Non-linear Prediction.
\newblock  Accepted for the European Signal Processing Conference (EUSIPCO).
  European Association for Signal Processing (EURASIP),  2021.

\bibitem[Omidvarnia \em{et~al.}(2013)Omidvarnia, Azemi, Boashash, O’Toole,
  Colditz, and Vanhatalo]{omidvarnia2013measuring}
Omidvarnia, A.; Azemi, G.; Boashash, B.; O’Toole, J.M.; Colditz, P.B.;
  Vanhatalo, S.
\newblock Measuring time-varying information flow in scalp EEG signals:
  orthogonalized partial directed coherence.
\newblock {\em IEEE transactions on biomedical engineering} {\bf 2013}, {\em
  61},~680--693.

\bibitem[Cover and Thomas(2012)]{cover2012elements}
Cover, T.M.; Thomas, J.A.
\newblock {\em Elements of information theory}; John Wiley \& Sons,  2012.

\bibitem[Baboukani \em{et~al.}(2019)Baboukani, Azemi, Boashash, Colditz, and
  Omidvarnia]{baboukani2019novel}
Baboukani, P.S.; Azemi, G.; Boashash, B.; Colditz, P.; Omidvarnia, A.
\newblock A novel multivariate phase synchrony measure: Application to
  multichannel newborn EEG analysis.
\newblock {\em Digital Signal Processing} {\bf 2019}, {\em 84},~59--68.

\bibitem[Schreiber(2000)]{schreiber2000measuring}
Schreiber, T.
\newblock Measuring information transfer.
\newblock {\em Physical review letters} {\bf 2000}, {\em 85},~461.

\bibitem[Baboukani \em{et~al.}()Baboukani, Mohammadi, and
  Azemi]{baboukani2017classifying}
Baboukani, P.S.; Mohammadi, S.; Azemi, G.
\newblock Classifying Single-Trial EEG During Motor Imagery Using a
  Multivariate Mutual Information Based Phase Synchrony Measure.
\newblock  2017 24th National and 2nd International Iranian Conference on
  Biomedical Engineering (ICBME). IEEE, pp. 1--4.

\bibitem[Gençağa(2018)]{Gen2018}
Gençağa, D.
\newblock Transfer Entropy.
\newblock {\em Entropy} {\bf 2018}, {\em 20}.
\newblock
  doi:{\changeurlcolor{black}\href{https://doi.org/10.3390/e20040288}{\detokenize{10.3390/e20040288}}}.

\bibitem[Faes \em{et~al.}(2017)Faes, Marinazzo, and
  Stramaglia]{faes2017multiscale}
Faes, L.; Marinazzo, D.; Stramaglia, S.
\newblock Multiscale information decomposition: Exact computation for
  multivariate Gaussian processes.
\newblock {\em Entropy} {\bf 2017}, {\em 19},~408.

\bibitem[Derpich \em{et~al.}(2013)Derpich, Silva, and
  {\O}stergaard]{derpich2013fundamental}
Derpich, M.S.; Silva, E.I.; {\O}stergaard, J.
\newblock Fundamental inequalities and identities involving mutual and directed
  informations in closed-loop systems.
\newblock {\em arXiv preprint arXiv:1301.6427} {\bf 2013}.

\bibitem[Massey(1990)]{massey1990causality}
Massey, J.
\newblock Causality, feedback and directed information.
\newblock  Proc. Int. Symp. Inf. Theory Applic.(ISITA-90). Citeseer,  1990, pp.
  303--305.

\bibitem[Wiener(1956)]{wiener1956theory}
Wiener, N.
\newblock The theory of prediction. Modern mathematics for engineers.
\newblock {\em New York} {\bf 1956}, pp. 165--190.

\bibitem[James \em{et~al.}(2016)James, Barnett, and
  Crutchfield]{james2016information}
James, R.G.; Barnett, N.; Crutchfield, J.P.
\newblock Information flows? A critique of transfer entropies.
\newblock {\em Physical review letters} {\bf 2016}, {\em 116},~238701.

\bibitem[Lizier and Prokopenko(2010)]{lizier2010differentiating}
Lizier, J.T.; Prokopenko, M.
\newblock Differentiating information transfer and causal effect.
\newblock {\em The European Physical Journal B} {\bf 2010}, {\em 73},~605--615.

\bibitem[Montalto \em{et~al.}(2014)Montalto, Faes, and
  Marinazzo]{montalto2014mute}
Montalto, A.; Faes, L.; Marinazzo, D.
\newblock MuTE: a MATLAB toolbox to compare established and novel estimators of
  the multivariate transfer entropy.
\newblock {\em PloS one} {\bf 2014}, {\em 9},~e109462.

\bibitem[Kraskov \em{et~al.}(2004)Kraskov, St{\"o}gbauer, and
  Grassberger]{kraskov2004estimating}
Kraskov, A.; St{\"o}gbauer, H.; Grassberger, P.
\newblock Estimating mutual information.
\newblock {\em Physical review E} {\bf 2004}, {\em 69},~066138.

\bibitem[Lindner \em{et~al.}(2011)Lindner, Vicente, Priesemann, and
  Wibral]{lindner2011trentool}
Lindner, M.; Vicente, R.; Priesemann, V.; Wibral, M.
\newblock TRENTOOL: A Matlab open source toolbox to analyse information flow in
  time series data with transfer entropy.
\newblock {\em BMC neuroscience} {\bf 2011}, {\em 12},~119.

\bibitem[Wibral \em{et~al.}(2013)Wibral, Pampu, Priesemann, Siebenh{\"u}hner,
  Seiwert, Lindner, Lizier, and Vicente]{wibral2013measuring}
Wibral, M.; Pampu, N.; Priesemann, V.; Siebenh{\"u}hner, F.; Seiwert, H.;
  Lindner, M.; Lizier, J.T.; Vicente, R.
\newblock Measuring information-transfer delays.
\newblock {\em PloS one} {\bf 2013}, {\em 8}.

\bibitem[Bossomaier \em{et~al.}(2016)Bossomaier, Barnett, Harr{\'e}, and
  Lizier]{bossomaier2016introduction}
Bossomaier, T.; Barnett, L.; Harr{\'e}, M.; Lizier, J.T.
\newblock An introduction to transfer entropy.
\newblock {\em Cham: Springer International Publishing} {\bf 2016}, pp. 65--95.

\bibitem[Ruiz-G{\'o}mez \em{et~al.}(2019)Ruiz-G{\'o}mez, Hornero, Poza,
  Maturana-Candelas, Pinto, and G{\'o}mez]{ruiz2019computational}
Ruiz-G{\'o}mez, S.J.; Hornero, R.; Poza, J.; Maturana-Candelas, A.; Pinto, N.;
  G{\'o}mez, C.
\newblock Computational modeling of the effects of EEG volume conduction on
  functional connectivity metrics. Application to Alzheimer’s disease
  continuum.
\newblock {\em Journal of neural engineering} {\bf 2019}, {\em 16},~066019.

\bibitem[Faes \em{et~al.}(2016)Faes, Marinazzo, Nollo, and
  Porta]{faes2016information}
Faes, L.; Marinazzo, D.; Nollo, G.; Porta, A.
\newblock An information-theoretic framework to map the spatiotemporal dynamics
  of the scalp electroencephalogram.
\newblock {\em IEEE Transactions on Biomedical Engineering} {\bf 2016}, {\em
  63},~2488--2496.

\bibitem[Mehta and Kliewer(2017)]{mehta2017directional}
Mehta, K.; Kliewer, J.
\newblock Directional and Causal Information Flow in EEG for Assessing
  Perceived Audio Quality.
\newblock {\em IEEE Transactions on Molecular, Biological and Multi-Scale
  Communications} {\bf 2017}, {\em 3},~150--165.

\bibitem[Zhang(2018)]{zhang2018low}
Zhang, J.
\newblock Low-dimensional approximation searching strategy for transfer entropy
  from non-uniform embedding.
\newblock {\em PloS one} {\bf 2018}, {\em 13},~e0194382.

\bibitem[Xiong \em{et~al.}(2017)Xiong, Faes, and Ivanov]{xiong2017entropy}
Xiong, W.; Faes, L.; Ivanov, P.C.
\newblock Entropy measures, entropy estimators, and their performance in
  quantifying complex dynamics: Effects of artifacts, nonstationarity, and
  long-range correlations.
\newblock {\em Physical Review E} {\bf 2017}, {\em 95},~062114.

\bibitem[Jia \em{et~al.}(2019)Jia, Lin, Jiao, Ma, and Wang]{jia2019detecting}
Jia, Z.; Lin, Y.; Jiao, Z.; Ma, Y.; Wang, J.
\newblock Detecting causality in multivariate time series via non-uniform
  embedding.
\newblock {\em Entropy} {\bf 2019}, {\em 21},~1233.

\bibitem[Kugiumtzis(2013)]{kugiumtzis2013direct}
Kugiumtzis, D.
\newblock Direct-coupling information measure from nonuniform embedding.
\newblock {\em Physical Review E} {\bf 2013}, {\em 87},~062918.

\bibitem[Olejarczyk \em{et~al.}(2017)Olejarczyk, Marzetti, Pizzella, and
  Zappasodi]{olejarczyk2017comparison}
Olejarczyk, E.; Marzetti, L.; Pizzella, V.; Zappasodi, F.
\newblock Comparison of connectivity analyses for resting state EEG data.
\newblock {\em Journal of neural engineering} {\bf 2017}, {\em 14},~036017.

\bibitem[Novelli \em{et~al.}(2019)Novelli, Wollstadt, Mediano, Wibral, and
  Lizier]{novelli2019large}
Novelli, L.; Wollstadt, P.; Mediano, P.; Wibral, M.; Lizier, J.T.
\newblock Large-scale directed network inference with multivariate transfer
  entropy and hierarchical statistical testing.
\newblock {\em Network Neuroscience} {\bf 2019}, {\em 3},~827--847.

\bibitem[May \em{et~al.}(2008)May, Maier, Dandy, and Fernando]{may2008non}
May, R.J.; Maier, H.R.; Dandy, G.C.; Fernando, T.G.
\newblock Non-linear variable selection for artificial neural networks using
  partial mutual information.
\newblock {\em Environmental Modelling \& Software} {\bf 2008}, {\em
  23},~1312--1326.

\bibitem[Li \em{et~al.}(2015)Li, Maier, and Zecchin]{li2015improved}
Li, X.; Maier, H.R.; Zecchin, A.C.
\newblock Improved PMI-based input variable selection approach for artificial
  neural network and other data driven environmental and water resource models.
\newblock {\em Environmental Modelling \& Software} {\bf 2015}, {\em
  65},~15--29.

\bibitem[Altman(1992)]{altman1992introduction}
Altman, N.S.
\newblock An introduction to kernel and nearest-neighbor nonparametric
  regression.
\newblock {\em The American Statistician} {\bf 1992}, {\em 46},~175--185.

\bibitem[Faes \em{et~al.}(2013)Faes, Nollo, and Porta]{faes2013compensated}
Faes, L.; Nollo, G.; Porta, A.
\newblock Compensated transfer entropy as a tool for reliably estimating
  information transfer in physiological time series.
\newblock {\em Entropy} {\bf 2013}, {\em 15},~198--219.

\bibitem[Faes \em{et~al.}(2015)Faes, Kugiumtzis, Nollo, Jurysta, and
  Marinazzo]{faes2015estimating}
Faes, L.; Kugiumtzis, D.; Nollo, G.; Jurysta, F.; Marinazzo, D.
\newblock Estimating the decomposition of predictive information in
  multivariate systems.
\newblock {\em Physical Review E} {\bf 2015}, {\em 91},~032904.

\bibitem[Danafar \em{et~al.}(2014)Danafar, Fukumizu, and
  Gomez]{danafar2014kernel}
Danafar, S.; Fukumizu, K.; Gomez, F.
\newblock Kernel-based Information Criterion.
\newblock {\em arXiv preprint arXiv:1408.5810} {\bf 2014}.

\bibitem[Faes \em{et~al.}(2014)Faes, Marinazzo, Montalto, and
  Nollo]{faes2014lag}
Faes, L.; Marinazzo, D.; Montalto, A.; Nollo, G.
\newblock Lag-specific transfer entropy as a tool to assess cardiovascular and
  cardiorespiratory information transfer.
\newblock {\em IEEE Transactions on Biomedical Engineering} {\bf 2014}, {\em
  61},~2556--2568.

\bibitem[Kramer \em{et~al.}(2008)Kramer, Kolaczyk, and
  Kirsch]{kramer2008emergent}
Kramer, M.A.; Kolaczyk, E.D.; Kirsch, H.E.
\newblock Emergent network topology at seizure onset in humans.
\newblock {\em Epilepsy research} {\bf 2008}, {\em 79},~173--186.

\bibitem[Wibral \em{et~al.}(2014)Wibral, Vicente, and
  Lizier]{wibral2014directed}
Wibral, M.; Vicente, R.; Lizier, J.T.
\newblock {\em Directed information measures in neuroscience}; Springer,  2014.

\bibitem[Barnett \em{et~al.}(2009)Barnett, Barrett, and
  Seth]{barnett2009granger}
Barnett, L.; Barrett, A.B.; Seth, A.K.
\newblock Granger causality and transfer entropy are equivalent for Gaussian
  variables.
\newblock {\em Physical review letters} {\bf 2009}, {\em 103},~238701.

\bibitem[Cole \em{et~al.}(2010)Cole, Platt, Schisterman, Chu, Westreich,
  Richardson, and Poole]{cole2010illustrating}
Cole, S.R.; Platt, R.W.; Schisterman, E.F.; Chu, H.; Westreich, D.; Richardson,
  D.; Poole, C.
\newblock Illustrating bias due to conditioning on a collider.
\newblock {\em International journal of epidemiology} {\bf 2010}, {\em
  39},~417--420.

\bibitem[Williams and Beer(2011)]{williams2011generalized}
Williams, P.L.; Beer, R.D.
\newblock Generalized measures of information transfer.
\newblock {\em arXiv preprint arXiv:1102.1507} {\bf 2011}.

\end{thebibliography}



\end{document}